\documentclass[a4paper,fleqn]{cas-sc}
\usepackage[authoryear,round]{natbib} 
\usepackage{caption}
\usepackage{subcaption}
\usepackage{rotating}
\usepackage{tabularx}
\usepackage{blindtext}

\def\om{\omega}
\def\Om{\Omega}

\def\de{\partial}
\def\H{\mathcal{H}}
\def\Z{\mathcal{Z}}

\def\his{\frac{i_S}{2}}

\def\heta{\hat{\eta}}
\def\tsc#1{\csdef{#1}{\textsc{\lowercase{#1}}\xspace}}
\tsc{WGM}
\tsc{QE}
\tsc{EP}
\tsc{PMS}
\tsc{BEC}
\tsc{DE}

\begin{document}
\let\WriteBookmarks\relax
\def\floatpagepagefraction{1}
\def\textpagefraction{.001}

\shorttitle{Analytical and numerical estimates for SRP semi-secular resonances}

\shortauthors{R. Paoli}

\title[mode = title]{Analytical and Numerical Estimates for 
Solar Radiation Pressure Semi-secular Resonances}        
\tnotemark[0]

\tnotetext[0]{This research has been funded by the European Union’s Horizon 2020 research
and innovation programme under the Marie Skłodowska-Curie grant agreement No 813644, Stardust-R.}


%
\author{Roberto Paoli}{}



\ead{roberto.paoli@uaic.ro}



\address{Faculty of Mathematics, Universitatea Alexandru Ioan Cuza, Bulevardul Carol I 11,
Iasi,
    700506, 
Romania}

\begin{abstract}
The aim of this work is to provide new insights on the dynamics associated to the resonances which arise as a consequence of the coupling of the effect due to the oblateness of the Earth and the Solar Radiation Pressure (SRP) effect for an uncontrolled object with moderate to high area-to-mass ratio. Analytical estimates for the location of the resulting resonant equilibrium points are provided, together with formulas to compute the maximum amplitude of the corresponding variation in the eccentricity, as a function of the initial conditions of the object and of its \textit{area-to-mass} ratio. The period of the variations of the eccentricity and inclinations due to such resonances is estimated using classical formulas. A classification based on the strength of the SRP resonances is provided. The estimates presented in the paper are validated using numerical tools, including the use of Fast Lyapunov Indicators to draw phase portraits and bifurcation diagrams. Many FLI maps depicting the location and overlapping of SRP resonances are presented. The results from this paper suggest that SRP resonances could be modeled in the context of either the Extended Fundamental Model by \cite{breiter_aps} or the Second Fundamental Model by \cite{henlem}.

\end{abstract}

\maketitle

\section{Introduction}
The growth in space activities around the Earth has increased immensely during the last few decades, with many benefits to scientific research, public and private companies, and ordinary people. Nowadays, thousands of man-made objects surround the planet, in all the orbiting zones, for many different tasks. However, some fragments of these spacecraft may detach during the de-orbiting stages of missions, or due to catastrophic events, producing swarms of new orbiting bodies. The serious threat caused by these \textit{space debris} for future missions and for the whole space environment has been assessed in \cite{kessler} and \cite{klinkrad2006space}. Therefore, studying the evolution of the orbits of such objects has become of seminal importance both for assessing possible future dangerous scenarios to spacecraft and humans in space, and for designing appropriate active removal strategies.

The goal of this paper is to study the \textit{resonant motion} which arises as a consequence of the coupling of the Earth's oblateness, encoded in the $J_2$ term of the classical spherical harmonics expansion of the geopotential (see \cite{kaula_spher}), with the Solar Radiation Pressure (hereafter SRP) effect, for an uncontrolled object with large \textit{area-to-mass ratio}.
Resonant dynamics is explored within the framework of a secular model obtained by averaging over the mean motion.
The SRP effect is due to the absorption and reflection of photons on the surface of the debris and it is known to have an effect on the eccentricity of the object \cite{hug77}. We adopt the cannonball approximation, which is equivalent to assuming that the sunlight is always perpendicular to the surface of the debris, and we do not consider the effect of the Earth's shadow. The problem of the coupling of the $J_2$ and SRP effects has recently been studied in a number of works. \cite{COLOMBO2012137}, performed a parametric study using various values of the semi-major axis and area-to-mass ratio, in order to provide the location of the equilibrium points that appear as a consequence of the coupling. \cite{alecol19} provided a description of the phase space associated to the problem. \cite{Schettino} used frequency analysis techniques to study low Earth orbits, including the effect of SRP. \cite{valk08}, performed a detailed analysis of the dynamics of an object with very large area-to-mass ratio in the Geosynchronous–Earth–Orbits (hereafter GEO) region. 
This research aims to complement the results of these previous works by providing analytical formulas to approximate the location, amplitude and period of the resonances arising from the coupling of the $J_2$ and SRP effects, which are known to act on long timescales. 
\color{black}
\cite{hug77} identifies the six most relevant terms in the SRP potential expansion, all of which are of first-degree in the semi-major axis. The associated cosine arguments involve a linear combination of the \textit{argument of perigee} $\om$, the \textit{right ascension of the ascending node} $\Om$ and the \textit{mean anomaly of the Sun} $M_S$. In view of the presence of $M_S$ the associated resonances are called \textit{semi-secular} by \cite{celletti20}. Their effect can be studied individually using \textit{toy models} which include only one of the above mentioned terms. 
This is equivalent to assuming that it is possible to average out all the other resonant terms using a close to the identity canonical transformation. \color{black} We adopt the Hamiltonian formalism, since it allows us to study resonances efficiently and to provide many useful information such as the maximum variation in the orbital elements and to approximate the period of such long-term variations. In order to give a new insight on the problem, we are going to proceed as in \cite{breiter_aps}, since the techniques used to study lunisolar resonances can be easily adapted to the SRP ones. We provide estimates for the location of the resonant equilibrium points and for the amplitude of their islands of stability, also called \textit{resonant width}, together with a smallness parameter to assess when such approximations hold. We detect the "strongest" resonances by comparing their maximum resonant width.
The approximations are then validated and discussed using numerical methods for orbit propagation, and for computing the Fast Lyapunov Indicators, which can be used to create bifurcation diagrams and, ultimately, to provide a cartography of the phase space. We show how the phase space near a resonance generally resembles the one associated to a pendulum. However, for some large area-to-mass ratio objects and for some initial conditions one obtains two near couples of pendulum like equilibria that could \textit{overlap}, showing a radically different phase portrait from the one of the pendulum. This suggests that SRP resonances could be modeled in terms of the Extended Fundamental Model (EFM) of resonance defined in \cite{breiter_ext}, as an extension of the Second Fundamental Model (SFM) of resonance by \cite{henlem}.

This paper is structured as follows.
In Section \ref{model} we define the model problem and some first approximations, up to the development of six \textit{toy models}. In Section \ref{approxcha} we adapt the procedure by \cite{breiter_aps} to the SRP case with low $a$ and moderate $\frac{A}{m}$, justifying all relevant approximations and providing analytical estimates for the maximum resonant width.
Section \ref{FLIsec} is dedicated to the results of tests obtained using Fast Lyapunov Indicators (hereafetr FLIs), a dynamical indicator that allows us to distinguish between chaotic and regualar motion, and also to easily plot phase spaces and bifurcation diagrams. Finally, in Section \ref{concl} we describe the results of this research and some ideas about possible future works.
\section{Model Description}\label{model}
\begin{figure}
    \centering
    \includegraphics[width=.75\textwidth]{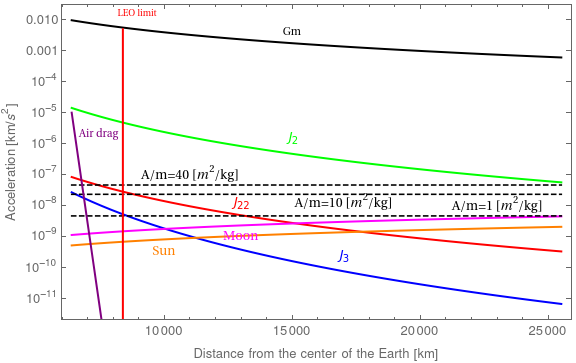}
    \caption{Comparison of the magnitude of the perturbations influencing the orbital motion of an uncontrolled object orbiting around the Earth. This picture is a reproduction of the one found in \cite{valk08}. If the area-to-mass ratio is large enough, the SRP effect (dashed line) is dominant over the lunisolar perturbation (orange and pink lines), at least in the upper LEO and lower MEO regions. The main geopotential term is given by $J_2$; in the following we will neglect the contribution of higher degree and order terms since their effect is not relevant over the timescales under consideration and/or in the region of phase space under investigation.}
    \label{magnitude}
\end{figure}

We study the dynamics of an object with moderate to high area-to-mass ratio in orbit around the Earth, in either the high Low–Earth–Orbits (hereafter LEO) region or the low Medium–Earth–Orbits (hereafter MEO) region. Figure \ref{magnitude} can be used to rank the perturbations affecting an object orbiting around the Earth. Therefore, we conclude that the most relevant perturbations affecting the Keplerian orbit are due to the geopotential, the lunisolar perturbation and, if the area-to-mass ratio $\frac{A}{m}$ is large enough, the Solar Radiation Pressure (SRP) effect. In order to give a mathematical insight to the problem, we adopt the Hamiltonian formalism. To do so, we use \textit{Delaunay elements} $(L,G,H,M,\omega,\Omega)$, where $L,G$ and $H$ are the \textit{actions} and the mean anomaly $M$, the argument of perigee $\omega$ and the right ascension of the ascending node $\Omega$ are their conjugated \textit{angles}.
To fix the notation, we recall the expression of the Delaunay elements $L,G$ and $H$ in terms of the semi-major axis $a$, the orbital eccentricity $e$ and the inclination $i$:
\begin{equation}
    L=\sqrt{\mu a}, \quad G=L\sqrt{1-e^2}, \quad H=G\cos i,
\end{equation}
where $\mu$ is the gravitational parameter of the Earth, defined as $\mathcal{G} \mathcal{M}_E$, $\mathcal{G}$ is the gravitational constant and $\mathcal{M}_E$ is the mass of the Earth. The Hamiltonian function is given by
\begin{equation}\label{hamfull}
    \mathcal{H}(L,G,H,M,\om,\Om)= -\frac{\mu}{2 L^2} + \mathcal{H}_{geo}+ \H_S + \H_M + \H_{SRP},
\end{equation}
where the first term is the two-body Keplerian energy, $\H_{geo}$ is the perturbation due to the geopotential, $\H_S$ and $\H_M$ are the lunisolar perturbations and $\H_{SRP}$ is the perturbation due to the SRP effect. The complete series expansion of all these perturbations is provided in Appendix \ref{appA}.
In order to describe the long term effect of these perturbations, we will consider a model which is averaged over the mean anomaly $M$. As a consequence, the momentum $L$ is a constant of motion of the problem and therefore also the (mean) semi-major axis is constant. Since the two-body Keplerian energy $-\frac{\mu}{2L^2}$ is a constant term, it can be neglected. Moreover, we retain only the most relevant terms in the expansions, i.e. the ones whose coefficients are the largest in magnitude. As a result, the model includes the $J_2$ secular terms of the geopotential\footnote{One could include terms of higher degree in the geopotential expansion while modeling the problem, but such contributions are small and short periodic and thus they can be neglected if the object is far from their associated resonances, such as the ones in the GEO and GPS regions (for the 1:1 and 2:1 resonances, respectively, involving $J_{22}$) and at the critical inclination resonance (involving $J_3$).}, and the dominant terms in the lunisolar and SRP potential expansions.
The lunisolar perturbations  expansions include both secular and non-secular terms of second degree in $a$. \cite{hug77} identifies the six most relevant terms in the SRP expansion, which are first degree terms in $a$ that depend linearly on the area-to-mass ratio, while \cite{alecol19} provide a description of the phase space 
of the resonant effect associated to such terms\color{black}. If the area-to-mass is large enough and the orbit is low, one may consider the SRP terms and the disturbing function due to the Earth as dominant and disregard the lunisolar perturbations. 
The resulting model Hamiltonian, which once again we label as $\H$ to keep a simple notation, is given by:
\begin{equation}
    \H= \H_{J_2} + \H_{SRP}.
\end{equation}
Let us first focus on the main perturbation to the Kepler problem. $\H_{J_2}$ is the Hamiltonian associated to the secular contribution due to the $J_2$ harmonic coefficient
\begin{equation}\label{hamj2}
    \mathcal{H}_{J_2}= \dfrac{1}{4} \dfrac{J_2 R_E^2 \mu^\frac{5}{2}}{ a^\frac{3}{2} G^3} \left(1-\dfrac{3H^2}{G^2} \right).
\end{equation}
It is well known (see \cite{celletti17}) that the $J_2$ term induces a slow variation in the Delaunay angles. In particular, the rate of change of $\om$ and $\Om$, expressed in terms of $a,e$ and $i$, is given by:
\begin{align}
    \dot{\omega}\simeq &\  4.98\left(\dfrac{R_E}{a}\right)^{\frac{7}{2}}(1-e^2)^{-2}(5\cos^2 i -1)\ ^\circ/\text{day},\label{roc}\\
    \dot{\Omega}\simeq & -9.97\left(\dfrac{R_E}{a}\right)^{\frac{7}{2}}(1-e^2)^{-2}\cos i\  ^\circ/\text{day}.\label{roc2}
\end{align}
The $\H_{SRP}$ part is the sum of the six most relevant terms in the expansion of the SRP perturbation potential:
\begin{equation}
     \H_{SRP}=\sum_{j,k} \H_{SRP}^{j,k}=\sum_{j,k}C_{j,k}\cos \sigma_{j,k},
\end{equation}
where $\sigma_{j,k}=\om + j \Om + k M_S + k \om_S$ and
\begin{equation}
    C_{j,k}=-\dfrac{3}{2}C_r P_r \dfrac{A}{m} \ \dfrac{L^2}{\mu}\ e\   
    \mathcal{F}_{j,k}(i,i_S),
\end{equation}
where $C_r$ is the reflectivity coefficient, depending on the optical properties of the surface of the object, $P_r$ is the radiation pressure for an object located at $a_S=1$ au, $\frac{A}{m}$ is the area-to-mass ratio with $A$ the cross section of the object and $m$ its mass. The argument of the perihelion $\omega_S$ and the inclination of the apparent orbit of the Sun $i_S$ can be considered as constants equal to $282.94^\circ$ and $23^\circ 26^{'} 21.4062^{''}$, respectively, as in \cite{celletti20}. The functions $\mathcal{F}_{j,k}$ are the product of two of the \textit{Kaula inclination functions} first defined in \cite{kaula_spher}. More precisely, one has that
\begin{equation}\label{kif}
    \mathcal{F}_{j,k}(i,i_S)=\begin{cases}
    F_{1,j,-1}(i)\ F_{1,j,\frac{k+1}{2}}(i_S) \quad \text{if }j\geq 0,\\
    F_{1,|j|,1}(i)\ F_{1,|j|,\frac{1-k}{2}}(i_S)\quad \text{if }j< 0.
    \end{cases}
\end{equation}
The definition of the Kaula inclination functions can be found in Appendix \ref{appA}. Figure \ref{magni} shows the graphs of $\mathcal{F}_{j,k}(i,i_S)$ as a function of the inclination of the object.
The analytical expression of the inclination functions in terms of both the Keplerian and Delaunay elements $G$ and $H$ is presented in Table \ref{inctable}. As far as the indices $j$ and $k$ are concerned, they must respect the following conditions:
\begin{equation}\label{conditionsabc}
    j\in \{-1,0,1\},\quad  k \in \{-1,1\}.
\end{equation}
The Hamiltonian $\H$ has two degrees of freedom and it is non-autonomous, since the cosine arguments in $\H_{SRP}^{j,k}$ depend implicitly on time through the mean anomaly of the Sun, $M_S$. A complete SRP potential expansion can be found in Appendix \ref{appA}.
\begin{table}
\begin{center}
\begin{minipage}{\textwidth}
\begin{tabular}{@{}llll@{}}
\toprule
$(j,k)$ & $\mathcal{F}_{j,k}$  & $\Tilde{\mathcal{F}}_{j,k}$ & $\sigma_{j,k}$\\
\midrule
$(0,\pm1)$   & $\mp\cos\his \sin\his \sin i$  & $\mp\cos\his \sin\his \sqrt{1-\frac{H^2}{G^2}}$   & $\om \pm (M_S +\om_S)$  \\
$(1,-1)$    &  $\cos^2\his\cos^2\frac{i}{2} $  &   $\frac{1}{2}\cos^2\his\left(1+\frac{H}{G}\right)$ &  $\om +\Om - M_S -\om_S$  \\
$(1,1)$    &  $\sin^2\his\cos^2\frac{i}{2} $ &  $\frac{1}{2}\sin^2\his\left(1+\frac{H}{G}\right)$     &  $\om +\Om + M_S+\om_S$   \\
$(-1,-1)$   &  $\sin^2\his\sin^2\frac{i}{2} $   &  $\frac{1}{2}\sin^2\his\left(1-\frac{H}{G}\right)$    &   $\om -\Om - M_S -\om_S$ \\
$(-1,1)$   &  $\cos^2\his\sin^2\frac{i}{2} $   &   $\frac{1}{2}\cos^2\his\left(1-\frac{H}{G}\right)$   &  $\om -\Om + M_S +\om_S$ \\\bottomrule
\end{tabular}
\end{minipage}
\end{center}
\caption{Analytical expressions for the inclination functions of the six most relevant terms in the SRP expansion. The expression of these functions in terms of the Delaunay elements is labeled by $\widetilde{\mathcal{F}}_{j,k}$.}\label{inctable}
\end{table}

\begin{figure}
    \centering
    \includegraphics[width=0.6\textwidth]{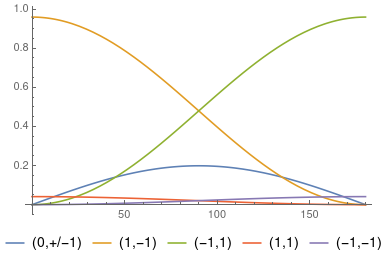}
    \caption{Plot of the inclination functions $\mathcal{F}_{j,k}(i,i_S)$ from Table \ref{inctable} as a function of the inclination for the indices $i$ and $j$ associated to the six most relevant terms in the expansion of the SRP perturbation potential.}
    \label{magni}
\end{figure}
The main goal of this research is to study the resonances due to the SRP effect. An ideal resonance occurs whenever the rate of change of the cosine argument of a term appearing in the Fourier expansion of a perturbation is equal to 0. Then the cosine argument, called the \textit{resonant angle}, is constant and the effect of the associated term accumulates over time. For example, in our scenario a resonance occurs whenever
\begin{equation}\label{reso}
    \dot{\om} + j\dot{\Om} + k n_S=0,
\end{equation}
where $n_S=\dot{M}_S$ is the mean motion of the Sun and the dot marks the derivative with respect to time. Resonances involving the Sun's or Moon's mean anomaly, such as the one defined by Eq. (\ref{reso}) are called \textit{semi-secular resonances} in \cite{celletti20}, while resonances involving the argument of perigee $\om$ are called \textit{apsidal resonances} in \cite{breiter_aps}. If Eq. (\ref{reso}) holds, the associated perturbative term, called the \textit{resonant term}, will have an \textit{enhanced effect} on the dynamics, which usually can be approximated using a simple pendulum. 
The approximate location of these resonances in the phase space, in the case of a small perturbation, can be found using Eqs. (\ref{roc}) and (\ref{roc2}), neglecting the contribution due to the resonant term. \cite{breiter_aps} uses this approach to compute the approximate location of second-degree lunisolar (semi-secular) apsidal resonances in the form of \textit{resonant curves}. Neglecting the resonant terms is equivalent to assuming that equilibria of different stability are realized for the same value of the eccentricity, thus simplifying the corresponding bifurcation sequences. The actual bifurcation plots, where the SRP contribution is not neglected would be slightly different depending on the value of $\sigma=0^\circ$ or $180^\circ$. In the following analysis we will focus on the \textit{toy models}\begin{equation}\label{toymodels}
\H_{j,k}(G,H,\om,\Om;L)=\H_{J_2}+\H_{SRP}^{j,k},
\end{equation}
which represent the situation in which all the SRP terms are averaged out except one. 
In fact, we validate this toy model by comparing it with the cartesian equations of motion that include all perturbing effects
\color{black}
The case of the overlapping of two distinct SRP resonant terms is briefly discussed numerically using Fast Laypunov Indicators at the end of Section \ref{FLIsec}.

\section{Analysis of SRP semi-secular resonances}\label{approxcha}
The goal of this section is to estimate the location of the six most relevant SRP semi-secular resonances. We will show that the approach presented in \cite{breiter_aps} can be adapted to the SRP case, with some differences, under a smallness condition which will defined later in the Section. This Section is structured as follows. We start by describing the procedure to study the effect of a specific resonant term, then we show the change of variables necessary to reduce the number of degrees of freedom to 1 and finally we approximate the equation to compute the equilibrium points of a given toy model. In particular, we describe how the dynamics qualitatively changes under the variation of parameters such as the inclination and the semi-major axis and we provide the equation of \textit{resonant curves} which allows us to estimate the location of the equilibrium solutions in terms of $a,e$ and $i$. We include an estimate of the maximum amplitude of the island of stability as a function of the semi-major axis, eccentricity and area-to-mass ratio.

We start by making the Hamiltonian $\H_{j,k}$ autonomous through the introduction of the \textit{dummy action} $\Lambda_S$, conjugated to $M_S$. In the following, we will express formulas in terms of the semi-major axis instead of the momentum $L$, since they both are constants of motion. The new Hamiltonian has three degree-of-freedom and can be written as
\begin{equation}
    \widetilde{\H}_{j,k}= n_S\  \Lambda_S + n \ a^2(\Z(G,H)+\mathcal{P}_{j,k}(\om,\Om,M_S,G,H)),
\end{equation}
where $n=\sqrt{\frac{\mu}{a^3}}$ and $n_S=\dot{M}_S$ are the satellite's and Sun's mean motions, respectively.

The function $\mathcal{Z}$ represent the secular part of the Earth's potential that was previously denoted by $\mathcal{H}_{J_2}$ and it is given by:
\begin{equation}
    \Z=\dfrac{3}{4}J_2\dfrac{R_E^2 n}{a^2 \eta^3}\left(\dfrac{1}{3}-c^2\right),
\end{equation}
where
\begin{equation}
    \eta=\sqrt{1-e^2}=\dfrac{G}{n \ a^2}, \quad \text{ and} \quad  c=\cos{i}=\dfrac{H}{G}.
\end{equation}
Moreover
\begin{equation}
    \mathcal{P}_{j,k}= C_{j,k}\cos{(\om + j \Om + k M_S+k \om_S)}, 
\end{equation}
where 
\begin{equation}
    C_{j,k}= -\dfrac{3}{2}  \dfrac{C_r P_r}{n\ a}\dfrac{A}{m} \ \sqrt{1-\eta^2} \mathcal{F}_{j,k}(i,i_S).
\end{equation}
Now we reduce the number of degrees-of-freedom of the Hamiltonian problem by means of a \textit{canonical transformation}.
For any pair of indices $(j,k)$ we can perform the following canonical transformation $\kappa_{a}$, which depends on the integral of motion $a$ 
\begin{equation}
    \kappa_{a}^{j,k}: (\om,\Omega,M_S,G,H,\Lambda_S; \H_{j,k}) \longrightarrow (\sigma_{j,k},\psi_{j,k},\chi_{j,k},\Phi_{j,k},\Psi_{j,k}, X_{j,k};\H_{j,k}^*),
\end{equation}
where
\begin{align}
    &\sigma_{j,k} = \om + j\Om + k M_S +k \om_S,\label{first}\\
    &\Phi_{j,k}= \dfrac{G}{\sqrt{\mu a}}= \eta, \\
    &\psi_{j,k}= \Om, \\
    &\Psi_{j,k}= \dfrac{H-j G}{\sqrt{\mu a}}=\eta(c-j),  \\
    &\chi_{j,k}= M_S,\\
    &X_{j,k}= \dfrac{\Lambda_S - k G}{\sqrt{\mu a}}.\label{last}
\end{align}
Transformations of the form given by formulas (\ref{first}) -- (\ref{last}) are not gauge-free, so the new Hamiltonian has to be divided by the transformation's valence $\sqrt{\mu a}$, resulting in
\begin{equation}\label{hetasigjk}
    \H_{j,k}^*= n_S\  X_{j,k}+ k\  n_S\  \eta + \Z+C_{j,k}\cos{\sigma_{j,k}}.
\end{equation}
Note that the new momenta are dimensionless quantities because of the normalization by the valence $\sqrt{\mu a}$.
In the following we are going to use the symbol $\eta$ instead of the formal $\Phi_{j,k}$. From (\ref{hetasigjk}) we deduce that $\psi_{j,k}\equiv \Om$ and $\chi_{j,k}\equiv M_S$ are cyclic variables, hence $\Psi_{j,k}$ and $X_{j,k}$ are constants of motion. Thus, the term $n_S \ X_{j,k}$ can be dropped from Equation (\ref{hetasigjk}), resulting in the reduced Hamiltonian
\begin{equation}\label{kjk}
    \mathcal{K}_{j,k}=\H_{j,k}^*-n_S X_{j,k}=k\  n_S\  \eta + \Z+C_{j,k}\cos{\sigma_{j,k}}.
\end{equation}
Let us also introduce a derived constant of motion
\begin{equation}\label{relationimp}
    \alpha_j\equiv-q_j \Psi_{j,k}=\eta(\lvert j\rvert - q_j c),
\end{equation}
where
\begin{equation}
    q_j=\begin{cases}
    1 \quad \text{if } j> 0\\
    -1 \quad \text{if } j\leq0.
    \end{cases}
\end{equation}
We remark that
\begin{equation}
    \alpha_0 \in (-1,1), \quad \alpha_{\pm 1} \in (0,2),
\end{equation}
since the inclination $i\in (0^\circ,180^\circ)$.
The Hamiltonian function $\mathcal{K}_{j,k}$ has one degree-of-freedom and it can be used to study the evolution of $(\sigma_{j,k},\eta)$. One simply needs to fix the parameters by using the initial condition of the considered spacecraft/debris and compute the value of the required integrals of motion. 
The expression of the inclination functions $\mathcal{F}_{j,k}(i,i_S)$ appearing in the SRP part of the Hamiltonian, as well as the inclination term in $\Z$, can be easily rewritten in terms of the integral of motion $\alpha_j$ using the relations
\begin{equation}
    c= q_j\left(\lvert j\rvert-\dfrac{\alpha_j}{\eta}\right), \quad s=\sqrt{1-c^2}.
\end{equation}
The expression of $\Z$ in terms of the constant $\alpha_j$ is given by
\begin{equation}
    \Z=-\dfrac{3}{4}J_2\dfrac{R_E^2 \ n}{a^2 \eta^5}\left[\alpha_j^2 -2\lvert j\rvert\alpha_j \eta -\eta^2\left(\dfrac{1}{3}-\lvert j \rvert^2\right) \right].
\end{equation}
For future reference, we provide also the first and second derivatives with respect to $\eta$:
\begin{align}
    &\Z':=\dfrac{\de\Z }{\de \eta}=\dfrac{3}{4}J_2\dfrac{R_E^2 \ n}{a^2 \eta^6}\left[5\alpha_j^2 -8\lvert j\rvert\alpha_j \eta -\eta^2(1-3 j^2) \right],\label{zprime}\\
    &\Z'':=\dfrac{\de^2\Z }{\de \eta^2}=-\dfrac{3}{2}J_2\dfrac{R_E^2 \ n}{a^2 \eta^7}\left[15\alpha_j^2 -20\lvert j\rvert\alpha_j \eta -2\eta^2(1-3 j^2) \right].
\end{align}
As far as the SRP term is concerned we distinguish between two cases: $j=0$ and $|j|=1$.
If $j=0$ one has that
\begin{equation}
    C_{0,k}=\dfrac{3}{2}\ k \   \dfrac{C_r P_r}{n\ a}\dfrac{A}{m}\sqrt{1-\eta^2} \ \cos\his \sin\his \sqrt{1-\left(\dfrac{\alpha_0}{\eta}\right)^2},
\end{equation}
while, if $|j|=1$,
\begin{equation}
    C_{\pm1,k}=-\dfrac{3}{2}  \dfrac{C_r P_r}{n\ a}\dfrac{A}{m}  \sqrt{1-\eta^2} \ cs(j,k;i_S)\  \left(1-\dfrac{\alpha_{\pm 1}}{2\eta}\right),
\end{equation}
where
\begin{equation}
    cs(j,k;i_S):=\begin{cases}\sin^2 \frac{i_S}{2}\quad \text{when } j\cdot k=1,\\
        \cos^2 \frac{i_S}{2} \quad \text{when } j\cdot k=-1.\end{cases}
\end{equation}
\begin{figure}
    \centering
    \includegraphics[height=.5\textwidth]{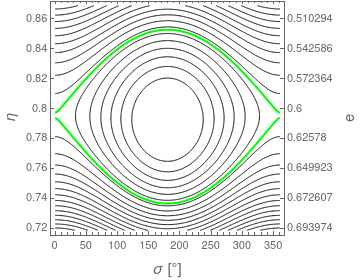}
    \caption{\footnotesize Example of a phase space exhibiting a couple of pendulum-like resonant equilibrium points, obtained using the $\mathcal{K}_{1,-1}$ toy model, with $a=13458.4$ km, $\alpha_1=0.1$, which corresponds to an inclination $i$ of $25.8419^\circ$ at $e=0$, and $\frac{A}{m}=1 \text{ m}^2$/kg. This picture has been obtained by plotting the level sets of the one degree-of-freedom reduced Hamiltonian $\mathcal{K}^*$. In particular the green curve correspond to the separatrix of the resulting pendulum-like phase space. One can appreciate a total variation in the eccentricity of $e_{max}-e_{min}\simeq0.16$, corresponding to a total variation in the perigee of roughly 2150 km.}
    \label{PLESRP}
\end{figure}
One can obtain the resulting phase space in the $(\eta,\sigma)$ plane by simply fixing the constant terms and plotting the level sets of the Hamiltonian $\mathcal{K}$, see Figure \ref{PLESRP}.

\subsection{Location of critical points, stability analysis and estimate of the resonance width}
In this section we adapt the procedure originally presented in \cite{breiter_aps} to approximate the location of the second-degree lunisolar semi-secular resonant equilibria, to the SRP case.
Moreover, we will address the stability of the SRP resonant equilibrium points and provide estimates for the amplitude of the islands of stability that surround the stable equilibrium points.

\begin{figure}
    \centering
    \includegraphics[height=.4 \textwidth]{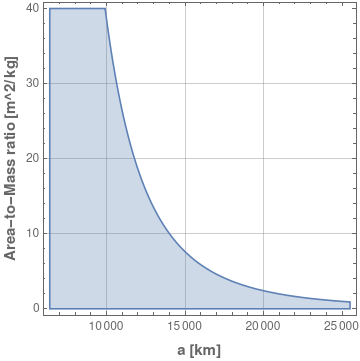} \includegraphics[height=.4 \textwidth]{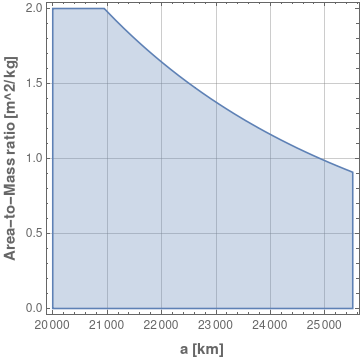}
    \caption{\footnotesize The highlighted region of the $\left(a,\frac{A}{m}\right)$ plane corresponds to the values which realize $\epsilon < 0.1$ (see Eq. (\ref{epsil})) and which allow us to approximate Eq. (\ref{soleta}) with Eq. (\ref{approxeta}); (left) from $a=1 R_E$ to $a=4 R_E$; (right) zoom of the bottom right corner of the plot on the left. Please note that in the LEO region this approximation is valid up to very large values of the area-to-mass ratio. Nonetheless, in the lower MEO region, one can consider moderate-to-high values of the area-to-mass ratio (up to $\simeq 1$ m\textsuperscript{2}/kg at $a=25000$ km).}
    \label{validity}
\end{figure}

We start by investigating the location of the equilibrium points of the reduced problem. The canonical equations of motion are given by
\begin{align}
    \dot{\sigma}_{j,k}= \dfrac{\de \mathcal{K}_{j,k}}{\de \eta}=&\  k\ n_S + \Z' + C'_{j,k} \cos{\sigma_{j,k}},\\
    \dot{\eta}=-\dfrac{\de \mathcal{K}_{j,k}}{\de \sigma_{j,k}}=&-C_{j,k} \sin{\sigma_{j,k}}, \label{doteta}
\end{align}
where the primes indicate partial derivatives with respect to $\eta$. From (\ref{doteta}) one can easily conclude that critical points with $\dot{\eta}=0$ exist at
\begin{enumerate}
    \item[A)] $\sigma_{j,k}=0^\circ$,
    \item[B)]$\sigma_{j,k}=180^\circ$,
    \item[C)] $\eta =1$, in which case $C_{j,k}=0$, regardless of the value of $\sigma_{j,k}$, or
    \item[D)] $F_{j,k}(\eta,\alpha_j,i_S)=0$, where again $C_{j,k}=0$.
\end{enumerate}
The two latter cases take place on the boundary of the admissible region, where $e=0$ and $i=0^\circ$ or $i=180^\circ$, but they should not be studied in the current action-angle cylindric parametrisation, because of the presence of \textit{virtual singularities}, see \cite{virtual}. Therefore, in the following we will focus only on points A or B, postponing the study of the above cases to a future work. 
 $F_{j,k}(\eta,\alpha_j,i_S)$ corresponds to the Kaula inclination function $\mathcal{F}_{j,k}(i,i_S)$ introduced in Eq. (\ref{kif}) expressed in terms of the new variables and constants of motion. The value of $\eta$ for the critical points A and B can be derived from the resonance condition $\dot{\sigma}_{j,k}=0$, i.e.
\begin{equation}\label{soleta}
    \dfrac{\de \mathcal{K}_{j,k}}{\de \eta}= k\ n_S + \Z' + \gamma C'_{j,k}=0.
\end{equation}
The symbol $\gamma$ selects the proper sign for a given critical point: $\gamma=1$ for point A, and $\gamma=-1$ for point B.  
Let us now suppose that we are sufficiently far from $\Z'=0$ where Eq. (\ref{soleta}) has no solutions. This condition is realized by the \textit{critical resonant inclinations} which induce a secular variation in the orbital elements due to the lunisolar perturbations or high terms of the geopotential. \color{black}
We wish to assess for which values of the semi-major axis and of the area-to-mass ratio one can neglect the contribution due to SRP from Eq. (\ref{soleta}). For this purpose, we define the following \textit{smallness parameter}
\begin{equation}\label{epsil}
    \epsilon := \dfrac{C_r P_r a}{J_2 R_E^2 n^2}\dfrac{A}{m}.
\end{equation}
If this parameter is indeed small, for example if $\epsilon<0.1$, the term $\Z'$ dominates over the SRP contribution $C'_{j,k}$, allowing us to neglect the latter. This reduces the validity of the subsequent approximation to the grey region depicted in Figure \ref{validity}. Nonetheless, we remark that in this approximation one can consider very high values of the area-to-mass ratio in LEO, and moderate values (larger than 0.5 m\textsuperscript{2}/kg) up to the GPS region, at around $a=25000$ km.
Neglecting the SRP contribution from Equation (\ref{soleta}), one obtains its approximate form, given by
\begin{equation}\label{approxeta}
    k n_S + \Z'=0.
\end{equation}
This is equivalent to assuming that the A and B points have the same value of $\eta$, therefore simplifying the bifurcation sequences. As we shall see from the numerical simulations of Section \ref{FLIsec}, this approximation 
is valid provided that $\epsilon$ is sufficiently small, for instance $\epsilon<0.1$.\color{black} However, we remark that by neglecting the SRP contribution we are disregarding a stable equilibrium point, located at either $\sigma=0^\circ$ or $180^\circ$, with a value of $\eta$ close to $1$. The location of this point is shown in \cite{alecol19} and it is detected numerically using Fast Lyapunov Indicators in Section \ref{FLIsec}.
The critical values of $\eta$ obtained as roots of the approximate Equation (\ref{approxeta}) are labeled $\heta$. By substituting (\ref{zprime}) into Equation (\ref{approxeta}) we can solve explicitly for $\alpha_j$, thus obtaining the following two roots
\begin{equation}\label{aljkl}
    \alpha_{j,k,\ell}= \dfrac{1}{5}\heta \left[4\lvert j \rvert + \ell \sqrt{5(1-k z_p \heta^4)+j^2}\right], \quad \ell=\pm 1,
\end{equation}
where
\begin{equation}
    z_p=\dfrac{4}{3}\dfrac{a^2\ n_S}{J_2 R_E^2 n}
\end{equation}
is the same auxiliary dimensionless parameter as the one defined in \cite{breiter_aps}. 
The formulas for $\alpha_{j,k,\ell}$ are slightly different from their counterparts in \cite{breiter_aps}, since they involve only first degree terms. \color{black}
In view of formula (\ref{aljkl}), one expects the resonances involving $\sigma_{j,k}$ and $\sigma_{-j,k}$ to be governed by similar expressions. However, notice that $\alpha_j$ and $\alpha_{-j}$ correspond to different inclinations for the same value of $\eta$:
\begin{equation}\label{cosrel}
    c(\alpha_{j})=-c(\alpha_{-j}),
\end{equation}
i.e. the cosines of the corresponding inclinations have opposite sign.

Formula (\ref{aljkl}) can be rewritten so to yield the location of the equilibria in the $(i,e)$ plane. More precisely one can derive a function $e_{eq}(i;j,k)$ so that for all values of $i$ in the domain of $e_{eq}(i;j,k)$, the couple $(i,e_{eq}(i;j,k))$ corresponds to an equilibrium solution: 
\begin{equation}\label{erescur}
    e_{eq}(i;j,k)=\left(1-\frac{1}{2}\sqrt{\dfrac{3J_2 n R_E^2}{k a^2 n_S }(1+2j\cos{i}-5\cos^2 i)}\right)^{1/2}.
\end{equation}
We notice that $e_{eq}(i;j,k)$ includes both branches of the resonant curves in one expression, so there is no need to use the label $l$.
One can also obtain an expression for the resonant curves as a function of the eccentricity:
\begin{equation}\label{irescur}
    i_{eq}(e;j,k,l)=\arccos\left[\frac{q_j}{5}\left(|j|-l\sqrt{5+j^2 -k\frac{20}{3}\frac{a^2 n_S}{J_2 R_E^2 n}(1-2e^2+e^4)}\ \right)\right].
\end{equation}
Similarly to the case of $e_{eq}$ shown above, one has that for all values of $e$ in the domain of $i_{eq}(e;j,k,l)$, the couple $(i_{eq}(e;j,k,l),e)$ corresponds to an equilibrium solution.
Notice that in this case one needs to label the function $i_{eq}$ using $l$, in order to distinguish between the different branches of the resonant curve.
\color{black}
In order to study the stability of the approximated critical points we consider the variational equation associated to the problem, obtained by linearising the equations of motion near an equilibrium point. The eigenvalues of the resulting matrix allow us to inspect the stability of an equilibrium point. They are given as solution of 
\begin{equation}
    \lambda^2 - 4 \gamma [C_{j,k}(\Z''+\gamma C_{j,k}'')]_{\eta=\heta}=0.
\end{equation}
If one is sufficiently far from \textit{the degenerate line} $\mathcal{Z}''=0$, for sufficiently low orbits and for small values of the area to mass ratio, the above equation can be approximated by
\begin{equation}\label{approxeigen}
    \lambda^2-4\gamma(C_{j,k}\Z'')_{\eta=\heta}=0.
\end{equation}
In order to approximate the equilibrium with $\heta$ we are assuming that we are far from $\Z'=0$, i.e. far from the critical inclination. From Equation (\ref{approxeigen}) it follows immediately that the points A and B have opposite indices of stability. Thus we need only to study the stability of A points to get all the relevant information; as an example purely imaginary values will indicate that A is stable and that B is unstable.

Let us now focus on the islands of stability surrounding a stable equilibrium point, i.e. on the area enclosed by the two branches of the separatrix. The amplitude of the island of stability, also called the \textit{resonance} (or \textit{separatrix}) \textit{width}, can be interpreted as a measure of the strength of a given resonance. Now we provide formulas to compute the resonant width as a function of the semi-major axis $a$, the integrals $\alpha_{j}$, the approximated equilibrium points $\heta$ and the area-to-mass ratio $\frac{A}{m}$. Assuming that the semi-amplitude of the resonance, $\Delta_{j,k}$, is a small quantity, one can use Taylor series to expand $\mathcal{K}(\eta +\Delta_{j,k},\hat{\sigma}_{j,k}+180^\circ)$ around $\hat{\eta}$ up to the second order, where $\hat{\sigma}_{j,k}$ is the value of the resonant angle corresponding to an unstable critical point. Then, using
\begin{equation}
    \mathcal{K}(\heta,\hat{\sigma}_{j,k})=\mathcal{K}(\heta+\Delta_{jk},\hat{\sigma}_{j,k}+180^\circ),
\end{equation}
and taking into account all previous considerations, we obtain that
\begin{equation}\label{amp}
    \Delta_{j,k}=2\left[\sqrt{\dfrac{\lvert C_{j,k}\rvert}{\lvert\Z''\rvert}}\right]_{\eta=\heta}.
\end{equation}
Approximating the Hamiltonian $\mathcal{K}$ around an equilibrium solution $\heta$ with
\begin{equation}\label{apppend}
    \overline{\mathcal{K}}(\eta, \sigma)=\dfrac{\mathcal{Z}^{''}(\heta)}{2}(\eta-\heta)^2 + C_{j,k}\left(\heta;\frac{A}{m}\right)\cos{\sigma},
\end{equation}
i.e. using a \textit{pendulum approximation} we are able to estimate the period of the circulating and librating solutions by using classical formulas for the pendulum (see for instance \cite{ferraz2007canonical}). Alternatively, one can compute the \textit{fundamental frequency at equilibrium}, which, in the case under consideration, is given by 
\begin{equation}\label{funfreq}
    \nu=\dfrac{\mathcal{Z}^{''}(\heta)}{4\pi}C_{j,k}\left(\heta;\frac{A}{m}\right),
\end{equation}
and use it to compute the \textit{fundamental period at the stable equilibrium}, given by
\begin{equation}\label{funper}
    T=\nu^{-1}.
\end{equation}
In the following we show that the above quantities will allow us to classify the SRP semi-secular resonances according to their amplitudes and periods. 
\color{black}
\subsection{Analysis of the toy models}

In this subsection we provide analytical approximations of the location of the equilibria and their associated separatrix width for the toy models $\mathcal{K}_{j,k}$. For each toy model we are going to qualitatively describe how the location of the resonant equilibria changes under the variation of the mean semi-major axis.

Expressions for all quantities of interest are provided in both the canonical variables $(\eta,\alpha)$ and the classical Keplerian orbital elements. We provide the expression of $\Delta_{j,k}$ in terms of the variable $\eta$ only. This choice stems from a practical reason: using the canonical variables, let $\eta_{eq}$ be a stable equilibrium solution and $\eta_{sep}^+$ and $\eta_{sep}^-$ the maximum and minum value assumed by $\eta$ along the two branches of the separatrix. One has that $|\eta_{eq}-\eta_{sep}^+|=|\eta_{eq}-\eta_{sep}^-|$. However, this symmetry is lost in the representation in terms of the eccentricity $e$. One can compute the corresponding quantities in terms of $e$ by simply using
\begin{align}
    e_{sep}^+&=\sqrt{1-(\eta_{sep}^-)^2}=\sqrt{1-(\eta_{eq}-\Delta)^2}\\
    e_{sep}^-&=\sqrt{1-(\eta_{sep}^+)^2}=\sqrt{1-(\eta_{eq}+\Delta)^2}.
\end{align}
The resonant curves depicting the location of the equilibria using both the Keplerian and canonical representations together with their resonance width are presented in Figures \ref{SRP0m1} through \ref{SRP11}. The blue lines in the plots correspond to the \textit{critical lines}, which represent the inclinations associated with the secular lunisolar resonances, while the orange ones correspond to the \textit{degenerate lines} along which the stability of the resonant equilibria changes. In the following section we will numerically investigate the phenomena of the \textit{merging} of resonances which take place in the neighborhood of a degenerate line.

Note that the resonance curves in the $(\eta,\alpha_j)$ plane correspond to the cases $m=0,\pm 2$ presented in \cite{breiter_aps}. However, SRP terms are of first order and they depend on the additional area-to-mass parameter, modifying the amplitude of the islands of stability. We can assume that the location of the equilibrium points does not change if the value of the area-to-mass ratio is small enough for the parameter $\epsilon$ to be small, i.e. when the $J_2$ term is dominant upon the SRP one. Larger values of the area to mass ratio actually change the location of the equilibrium points, as confirmed by the numerical tests in Section \ref{FLIsec}.

\begin{table}
\begin{center}
\begin{minipage}{\textwidth}
\begin{tabular}{@{}cccll@{}}
\toprule
$(j,k)$ & Key values & Semi-major axis  & Notes & Coordinates of the extremal points \\
 & of $z_p$ & [km] &\\
\midrule
$(0,-1)$  & 4& 15057.9 & Both branches of the resonant curve \\
& & & meet the boundaries $\alpha=\pm \eta$ \\
& & & \\
$(0,1)$    & 1/3& 7403.31 & Extremal points $E_{\ell}$ in the admissible region &  $\heta(E_{\ell})=(3z_p)^{-1/4}$ \\
    &  & & & $\alpha_0(E_{\ell})=\ell(\frac{675}{4}z_p)^{-1/4}$ \\
    & 1& 10133.2 & Junction point in the admissible region & \\
    & & & \\
$(\pm 1,-1)$   & $\frac{8\sqrt{10}-10}{45}$& 7445.06 & Maximum point of the lower branch, $E_{-1}$, & $\heta(E_{-1})=\left(\frac{8\sqrt{10}-10}{45 z_p}\right)^{1/4}$ \\
  & & & in the admissible region &  $\alpha_{\pm 1,-1}(E_{-1})=\heta(E_{-1})\frac{2}{15}(5-\sqrt{10})$\\
      & & &\\
    & 2& 12352.5 & Lower branch meets the boundary $\alpha=0$ & \\
    & & & & \\
    & 6 & 16907.3 & Upper branch meets the boundary $\alpha=2\eta$\\
    & & &\\
$(\pm 1,1)$   & $\frac{8\sqrt{10}+10}{45}$& 9453.98 & Maximum point of the upper branch, $E_{1}$,& $\heta(E_{1})=\left(\frac{8\sqrt{10}+10}{45 z_p}\right)^{1/4}$ \\
   &  & &  in the admissible region &  $\alpha_{\pm 1,-1,1}(E_{1})=\heta(E_{1})\frac{2}{15}(5+\sqrt{10})$\\
       & & \\
    & 6/5& 10675 & Junction point in the admissible region &\\\bottomrule
\end{tabular}
\end{minipage}
\end{center}
\caption{List of all possible key values, with a short description of how the curves qualitatively change in correspondence of each value. For more details, see \cite{breiter_aps}.}\label{zptable}
\end{table}
Proceeding as in \cite{breiter_aps}, we are going to identify \textit{key values} for the semi-major axis which separates two qualitatively different situations. In our analysis a key value will mark either the intersection of a resonant curve with the lines corresponding to $e=0$, $i=0^{\circ}$ and $i=180^{\circ}$ or the junction of two branches of the same resonant curves, or the appearance of an extremal point for the resonant curves in the $(\eta,\alpha_j)$ plane\footnote{
In the $(i,e)$ plane, this correspond to the case in which one branch of the resonant curve is tangent to one of the level sets of $\alpha_j$.}.
A list of all possible key values for the problem under consideration is presented in Table \ref{zptable}.
The qualitative description of the resonant curves is focused on the canonical representation in the $(\eta,\alpha_j)$ variables, since this is the direct procedure to analyse the problem. Moreover, we are going to highlight peculiar features of the same curves also in the $(i,e)$ plane, to provide a more insightful and practical perspective.

\begin{figure}[p]
 \centering
    \includegraphics[width=0.9\textwidth]{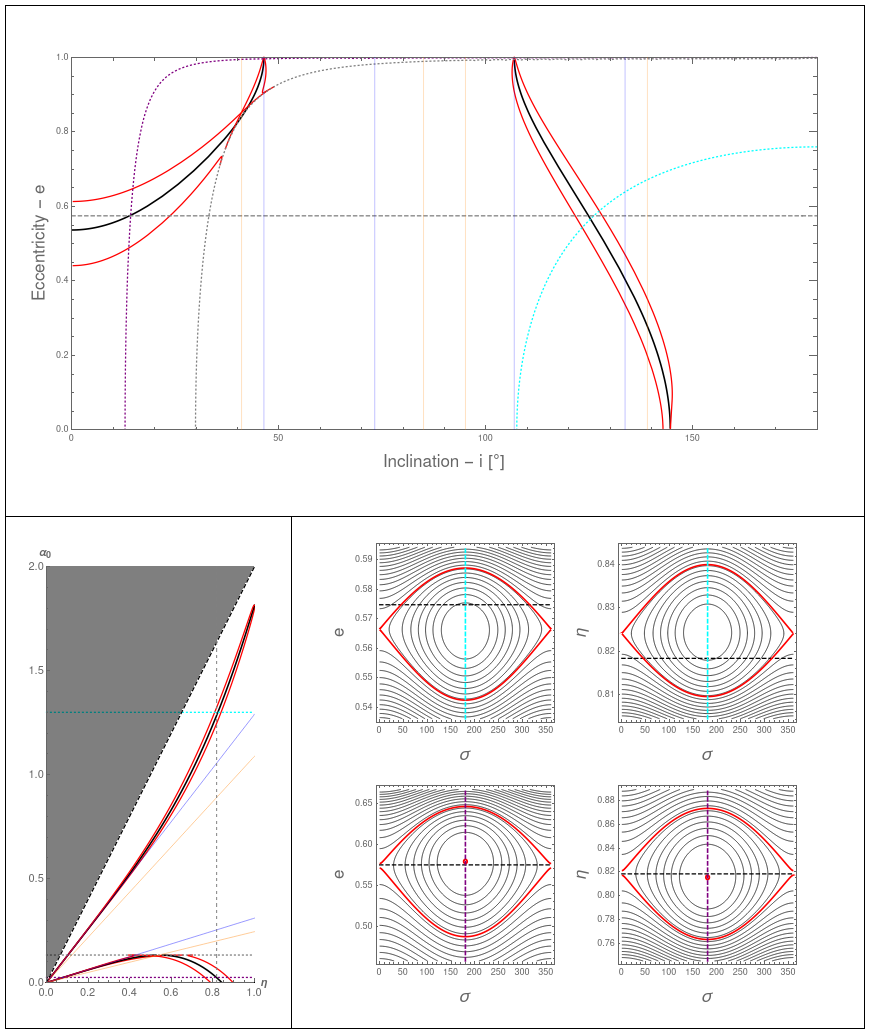}
    \caption{Resonance $(1,-1)$: Location of the equilibria and separatrix width for an object with $A/m=1$ m\textsuperscript{2}/kg, mean semi-major axis $a=15000$ km. The resonant curves are presented in the $(i,e)$ plane (\textit{top panel}) and in the $(\eta,\alpha)$ plane (\textit{bottom left panel}). The grey straight dashed lines mark the critical eccentricity which leads to re-entry. The dotted grey curve represents $\Z^{''}=0$ and it corresponds to a maximum of the resonant curve in the $(\eta,\alpha)$ plane. The corresponding curve in the $(i,e)$ plane is tangent to the resonant curves. The purple and cyan curves correspond to $\alpha=0.025$ and $\alpha=1.3$, respectively. The theoretical phase spaces for these two selected values of $\alpha$ are presented in the bottom row panel in terms of the resonant angle $\sigma$ and of the eccentricity $e$ (\textit{left colum}) and of the canonical variable $\eta$ (\textit{right column}). The red curves surrounding the resonant curves show the resonance width of the equilibria, which is plotted along the level sets of $\alpha$. These curves correspond to the intersection of the two branches of the separatrix with a straight line with $\sigma=0^{\circ},180^{\circ}$, depending on the resonance under consisderation.}
    \label{exampleres}
\end{figure}
As an example on how to interpret the Figures in this section, let us focus on Figure \ref{exampleres}, involving the $(1,-1)$ toy model for a debris with $A/m=1$ m\textsuperscript{2}/kg in the lower MEO region, at $a=15 000$ km. The black curves represent the resonant curves, i.e. the location of the equilibrium solutions, and they are presented in both the $(i,e)$ plane and the $(\eta,\alpha)$ plane. The red lines surrounding the equilibrium solutions represent
the resonance width and they are obtained by plotting the points of intersection between the separatrix and $\sigma=\sigma_{s}$ for each value of $\alpha$, where $\sigma_s$ correspond to a stable equilibrium point.
Also included in the plot are the critical inclinations (\textit{blue lines}), corresponding to secular lunisolar resonances, and the degenerate lines (\textit{orange lines}) which separate the resonant curves in two branches of opposite stability. Our theory is in principle not valid in the neighborhood of these lines. The straight grey dashed lines correspond to the critical eccentricity $e_{crit}$ which leads to re-entry. Only orbits with smaller eccentricity,  are considered admissible. On the $(\eta,\alpha)$ plane this is equivalent to considering only values of $\eta>\eta_{crit}=\eta(e_{crit})$. Moreover, when considering the canonical variables, only a triangular region of the plane is deemed admissible, due to Eq. (\ref{relationimp}). We recall that $\alpha$ is a constant of motion of the problem, so the dynamics takes place along the level sets $\alpha=\alpha(t_0)$, where $\alpha(t_0)$ depends on the initial conditions of the problem. Fixing $\alpha$ is equivalent to fixing a certain \textit{dynamical regime}, restricting the number of solutions and their associated variations in the eccentricity and inclination. Each intersection of $\alpha=\alpha(t_0)$ with a resonant curve corresponds to a \textit{Pendulum Like Equilibrium} (\textit{PLE}) solution. For example, if $\alpha=0.025$, corresponding to the purple curves of Figure \ref{exampleres}, there are three PLE. However, two of them exhibit a non-admissible value of the eccentricity, while the remaining one's equals the critical eccentricity. We remark that in this case there are no significative variations in the inclination, due to the shape of $\alpha=0.025$ in the $(i,e)$ plane. On the other hand, if we consider $\alpha=1.3$, corresponding to the cyan curves of Figure \ref{exampleres}, one can appreciate a more prominent variation of the inclination inside the resonant island. We notice that there are no other PLEs for this dynamical regime. On the bottom right panel we present the phase spaces in the neighborhood of two relevant PLE, in terms of both $(\sigma,e)$ and $(\sigma,\eta)$. Note that the symmetry of the separatrix width with respect to the stable equilibrium solution is lost in the $(\sigma,e)$ representation. We can now move to the qualitative description of the SRP semi-secular resonances.
\subsection*{Case $j=0$}
The associated resonant angles are given by $\sigma_{0,\pm 1}=g\pm M_S$.
If $k=-1$, we have that the resonant curves are bent upwards and symmetric with respect to the $\eta$-axis, i.e. symmetric to $i=90^{\circ}$ in the $(e,i)$ plane. The approximate resonant curves are defined by
\begin{equation}
    \alpha_{0,-1,l}=l\sqrt{\dfrac{1+ z_p \eta^4}{5}},
\end{equation}
or, equivalently,
\begin{equation}
    e_{eq}(i;0,-1)=\sqrt{1-\dfrac{1}{2 a}\sqrt{3 \dfrac{J_2\  n\  R_E^2}{n_s}(5 \cos^2 i -1) }}.
\end{equation}
\begin{figure}
 \centering
    \includegraphics[width=1\textwidth]{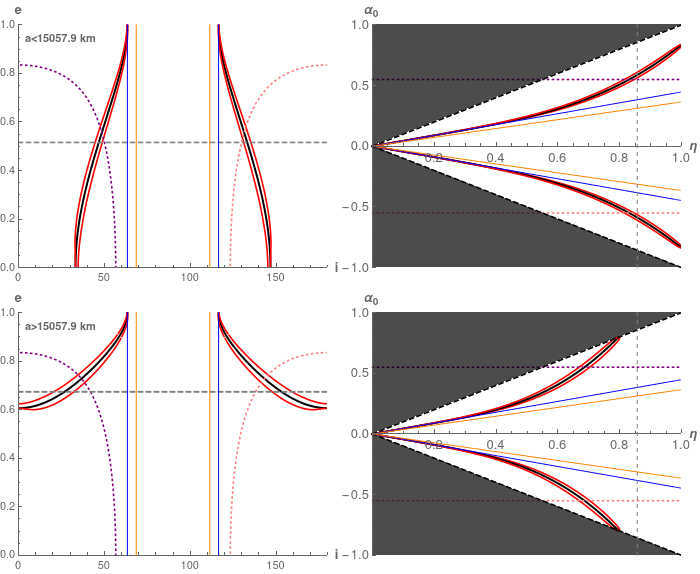} 
    \caption{Resonant curves (\textit{red lines}) for the resonance $j=0, k=-1$ in the $(i,e)$-plane (\textit{left column}) and in the $(\eta,\alpha)$-plane (\textit{right column}). The dark region in the plots on the right corresponds to a non-admissible region, see \cite{breiter_aps}. The plots in the top row depict the qualitative behaviour for $z_p\leq 4$, or $a\leq15057.9$ km, while the ones in the bottom row depict the qualitative behaviour for $z_p>4$, or $a>15057.9$ km. For reference, we included the \textit{critical lines} $\Z'=0$ (\textit{blue dashed lines}) and $\Z''$(\textit{blue dotted lines}). In both representations, the grey dashed line corresponds to the critical eccentricity which leads to a collision with the planet.}
    \label{SRP0m1}
\end{figure}

The only possible key value is given by $z_p=4$, corresponding to $a\simeq15057.9$. If $z_p<4$ the resonant curve crosses the $e=0$ line, while if $z_p>4$ the resonant curve meets the boundary of the admissible region, corresponding to either $i=0^{\circ},180^{\circ}$. However, the study of the behaviour of the solutions near these boundary regions require a different approach and is beyond the scope of this paper. The two cases are shown in Figure \ref{SRP0m1}, in both the $(i,e)$ and $(\eta,\alpha)$  planes. For all nonzero values of $\alpha_j$ we can have at most one pendulum-like couple of equilibrium points, which we call A and B as in the previous subsection. Using (\ref{approxeigen}) one has that A points are unstable, while B points are stable.
We observe that for small values of $z_p$, and thus of the semi-major axis $a$, the resonant curves in the $(i,e)$-plane are almost straight lines very close to the critical inclinations $i=63.4^\circ$ and $i=116.6^\circ$, while, for greater values of $a$ the resonant curves move closer to $e=1$. This is confirmed by the analytical expression of $e_{eq}(i;0,-1)$.

Using formula (\ref{amp}), the (semi-)amplitude of the resonant island surrounding a B point with coordinates $(\eta,180^\circ)$ is found to be
\begin{align}
    \Delta_{0,-1}=&\ 2\sqrt{\dfrac{A}{m}}\left(\dfrac{C_rP_r\ a^4}{J_2\  R_E^2\  \mu} \right)^{1/2} \eta^{5/2} (1-\eta^2)^{1/4}\left(\dfrac{c_s s_s}{1+3z_p\eta^4}\right)^{1/2}\lvert4-z_p \eta^4\rvert^{1/4}\\
    =&\  2\sqrt{\dfrac{A}{m}}\left(\dfrac{C_rP_r\ a^4}{J_2\  R_E^2\  \mu} \right)^{1/2} (1-e^2)^{5/4} \sqrt{e}\left(\dfrac{c_s s_s}{1+4\dfrac{a^2 n_S}{J_2 R_E^2 n}(1-e^2)^{2}}\right)^{1/2}\left\lvert4-\dfrac{4}{3}\dfrac{a^2 n_S}{J_2 R_E^2 n} (1-e^2)^2\right\rvert^{1/4} ,
\end{align}
where $c_s=\cos \frac{i_s}{2}$ and $s_s=\sin \frac{i_s}{2}$.
One can notice a common feature of SRP semi-secular resonances, namely the fact that the amplitude of the resonant islands grows as the square root of the area-to-mass ratio.

If $k=1$ the resonant curves are bent inwards and they might meet for $e>0$, depending on the value of $z_p$. The approximate resonant curves are defined by
\begin{equation}
    \alpha_{0,1,l}=l\sqrt{\dfrac{1-\ z_p \eta^4}{5}},
\end{equation}
or, equivalently,
\begin{equation}
    e_{eq}(i;0,1)=\sqrt{1-\dfrac{1}{2 a}\sqrt{3 \dfrac{J_2\  n\  R_E^2}{n_s}(1-5 \cos^2 i) }}.
\end{equation}

\begin{figure}

 \centering
    \includegraphics[width=1\textwidth]{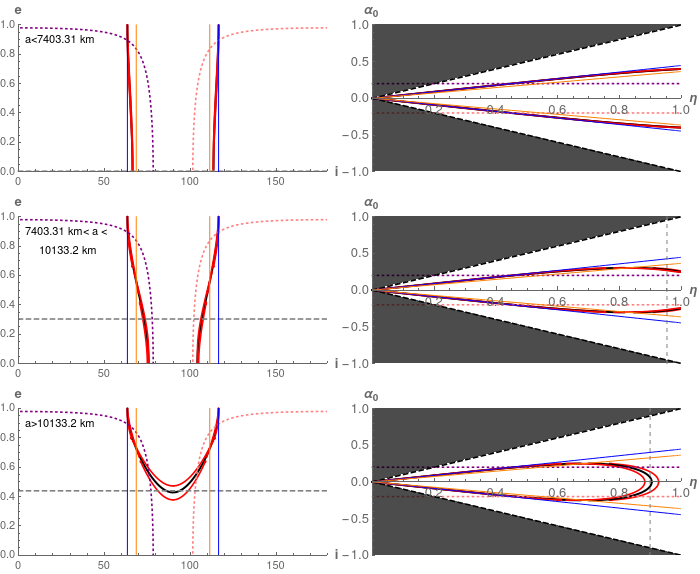} 
    \caption{Resonant curves for the resonance $j=0, k=1$ in the $(i,e)$-plane (left column) and in the $(\eta,\alpha)$-plane (right column). The choice of the colors and the represented curves are the same as in Figure \ref{SRP0m1}. The plots depict the three different qualitative scenarios for this resonance: $z_p\leq 1/3$, or $a\leq 7403.31$ km (\textit{top row}); $\frac{1}{3}<z_p\leq 1$, or $7403.31$ km $<a\leq 10133.2$ km (\textit{middle row}); $z_p>1$, or $a>10133.2$ km (\textit{bottom row}). Note that in the top row plots the critical eccentricity is very close to zero.}
    \label{SRP01}
\end{figure}

The two possible key values, corresponding to the appearance of the extremal points $E_{\pm 1}$ and of the junction point $J$ are presented in Table \ref{zptable}. The three possible scenarios are presented in Figure \ref{SRP01}. If $z_p>1/3$, i.e. if $a>7403.31$ km, up to two pairs of PLE can be found. However, in most cases only one of these two PLEs is a non colliding solution, i.e. the eccentricity is smaller than the critical one. Once again, for small values of $z_p$, the resonant curves are almost straight lines in the $(i,e)$-plane, which are very close to the critical inclinations. If $z_p>1$, i.e. if $a>10133.2$ km, the resonant curves meet at $i=90^\circ$ and they quickly get far from the $e=0$ line, apparently making these resonances not relevant for almost circular orbits in MEO. However, they could be very important for missions in highly-eccentric orbits and, as we shall see in Section \ref{FLIsec}, for high values of the area-to-mass ratio, the resonant islands could nonetheless get very close to $e=0$.
Following \cite{breiter_aps}, we mark the equilibrium to the right of $\eta(E_{\pm1})$ as $X_{\pm1,N}$ and the one to its left as $X_{\pm1,S}$, where $X$ is either $A$ or $B$. The notation is coherent even if $E_{\pm 1}$ is not in the admissible region. The stability of the equilibria is once again obtained by studying the sign of the solutions of (\ref{approxeigen}). One has that the stable equilibrium points are $A_{\pm1,N}$ and $B_{\pm 1,S}$.
The amplitude of the islands of stability surrounding the stable equilibrium points can again be found using (\ref{amp}), resulting in 

\begin{align}
    \Delta_{0,1}=&\ 2\sqrt{\dfrac{A}{m}}\left(\dfrac{C_rP_r\ a^4}{J_2\  R_E^2\  \mu} \right)^{1/2} \eta^{5/2} (1-\eta^2)^{1/4}\left(\dfrac{c_s s_s}{\lvert1-3z_p\eta^4\rvert}\right)^{1/2}(4+z_p \eta^4)^{1/4}\\
    =&\  2\sqrt{\dfrac{A}{m}}\left(\dfrac{C_rP_r\ a^4}{J_2\  R_E^2\  \mu} \right)^{1/2} (1-e^2)^{5/4} \sqrt{e}\left(\dfrac{c_s s_s}{\left\lvert 1-4\dfrac{a^2 n_S}{J_2 R_E^2 n}(1-e^2)^{2}\right\rvert}\right)^{1/2}\left(4+\dfrac{4}{3}\dfrac{a^2 n_S}{J_2 R_E^2 n} (1-e^2)^2\right)^{1/4}.
\end{align}
\subsection*{Case $j=\pm 1$}
Let us now focus on the cases $j=\pm 1$. They correspond to $m=\pm 2$ in \cite{breiter_aps}. Let us consider $j=1$. The other case can be obtained in a very simple way using the same quantities and relation (\ref{cosrel}).
The approximate resonant curves are defined by
\begin{equation}\label{resoj1}
    \alpha_{\pm 1,k,\ell}=\eta \left[\dfrac{4}{5}+\ell \sqrt{\dfrac{1}{5}\left(\dfrac{6}{5} - k z_p \eta^4 \right)} \right],
\end{equation}
or, equivalently,
\begin{equation}
    e_{eq}(i;\pm 1,k)=\sqrt{1-\dfrac{1}{2 a}\sqrt{3 \dfrac{J_2\  n\  R_E^2}{k\ n_s}(1\pm 2 \cos i-5 \cos^2 i) }}.
\end{equation}
 Note that in the canonical variables the resonant curves are the same for $j=1$ and $j=-1$, while the expression for $e_{eq}$ ultimately depends on the sign of $j$. 
If $k=-1$ the resonant curve has two branches bent outwards. However, unlike in the case $j=0$, the lower branch in the canonical representation can have a maximum point, labeled $E_{-1}$. This situation allows up to 3 couples of PLE for a fixed $\alpha_1$. 

\begin{figure}
\centering

    \includegraphics[width=1\textwidth]{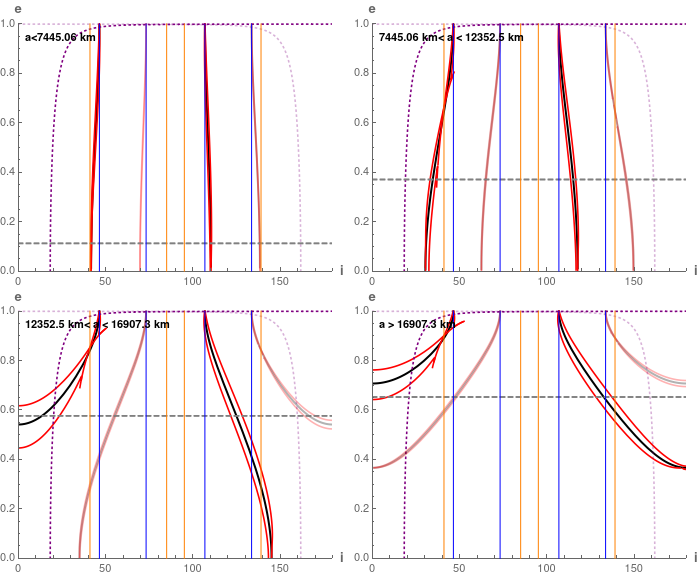} 
    \caption{Resonant curves for the resonance $j=\pm 1, k=-1$ in the $(i,e)$-plane. The plots depict the four different qualitative scenarios for this resonance: case $z_p\leq\frac{8\sqrt{10}-10}{45}$, or $a\leq 7445.06$ km (\textit{top left}); $\frac{8\sqrt{10}-10}{45}<z_p\leq 2$, or $7445.06$ km $<a\leq 12352.5$ km (\textit{top right}); $2<z_p\leq 6$, or $12352.5$ km $<a\leq 16907.3$ km (\textit{bottom left}); $z_p>6$, or $a>16907.3$ km (\textit{bottom right}). In the first panel the critical eccentricity is very low. The bold curves correspond to the $j=1$ case, while the opaque ones correspond to the $j=-1$ case. Note that the resonant curves corresponding to the different values of $j$ are symmetrical to $i=90^\circ$.}
    \label{SRP1m1eccinc}
\end{figure}

\begin{figure}
\centering

    \includegraphics[width=1\textwidth]{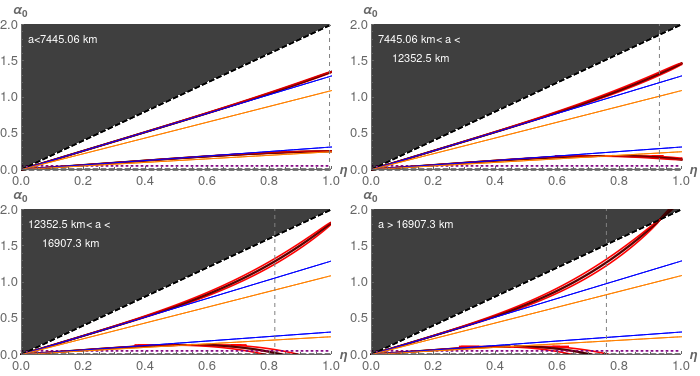} 
        \caption{Resonant curves in the $(\eta,\alpha)$-plane for the $j=\pm 1, k=-1$. This plots complement the ones from Figure \ref{SRP1m1eccinc}. In this representation there is no need to distinguish between the cases $j=1$ and $j=-1$.}
    \label{SRP1m1}
\end{figure}

Let us describe the key values of $z_p$ for the resonance $(1,-1)$. The first key value is given by $z_p=(8\sqrt{10}-10)/45$, that corresponds to $a=7445.06$ km . Indeed, if $z_p>(8\sqrt{10}-10)/45$ the maximum point of the lower branch, is such that $\eta(E_{-1})<1$, i.e. the maximum point exists for an orbit with nonzero eccentricity, resulting in a qualitative change of the resonant curve. The next key value is given by $z_p=2$, i.e. $a=12352.5$ km, which marks the intersection of the resonant curve defined by $\alpha_{1,-1,-1}$ with the boundary of the admissible region, i.e. the $\eta$ axis. Finally, the last key value is given by $z_p=6$, or $a=16907.3$ km, which marks the intersection of the upper branch of the resonant curve, $\alpha_{1,-1,1}$, with the boundary of the admissible region, given by $\alpha_1=2\eta$. The situation is depicted in Figures \ref{SRP1m1eccinc} and \ref{SRP1m1} in the two different representations. In Figure \ref{SRP1m1eccinc} we included the case $j=-1$ as opaque lines. The lower boundary of the admissible region in the canonical variables, $\alpha=0$, corresponds to $i=0^{\circ}$ if $j=1$ and to $i=180^{\circ}$ if $j=-1$, while they exchange in the case of the upper boundary $\alpha=2\eta$.

The stability of the equilibria is once again obtained by studying the sign of the solutions of (\ref{approxeigen}). One has that the stable equilibrium points are $A_{-1,S},B_{-1,N}$ and $B_{1}$.
The resonant width has two slightly different expressions depending on which branch of the resonant curve the equilibrium point belongs to: we will then include $l$ while labeling the resonant widths $\Delta_{1,-1,l}$.

\begin{align}
    \Delta_{1,-1,1}=&\ \sqrt{2}\sqrt{\dfrac{A}{m}}\left(\dfrac{C_rP_r\ a^4}{J_2\  R_E^2\  \mu} \right)^{1/2} \eta^{5/2} (1-\eta^2)^{1/4}\left(\dfrac{c_s^2\ \left\lvert 6-\sqrt{6+5 z_p \eta^4}\right\rvert}{6+15z_p\eta^4+4\sqrt{6+5z_p\eta^4}}\right)^{1/2}\\
    =&\  \sqrt{2}\sqrt{\dfrac{A}{m}}\left(\dfrac{C_rP_r\ a^4}{J_2\  R_E^2\  \mu} \right)^{1/2} (1-e^2)^{5/4} \sqrt{e}\left(\dfrac{c_s^2\ \left\lvert 6-\sqrt{6+\dfrac{20}{3}\dfrac{a^2 \ n_S}{J_2\  R_E^2\ n}(1-e^2)^2} \right\rvert}{6+20\dfrac{a^2 \ n_S}{J_2\  R_E^2\ n}(1-e^2)^2+4\sqrt{6+\dfrac{20}{3}\dfrac{a^2 \ n_S}{J_2\  R_E^2\ n}(1-e^2)^2}}\right)^{1/2} ,\\
     \Delta_{1,-1,-1}=&\ \sqrt{2}\sqrt{\dfrac{A}{m}}\left(\dfrac{C_rP_r\ a^4}{J_2\  R_E^2\  \mu} \right)^{1/2} \eta^{5/2} (1-\eta^2)^{1/4}\left(\dfrac{c_s^2\ \left( 6+\sqrt{6+5 z_p \eta^4}\right)}{\lvert6+15z_p\eta^4-4\sqrt{6+5z_p\eta^4}\rvert}\right)^{1/2}\\
    =&\  \sqrt{2}\sqrt{\dfrac{A}{m}}\left(\dfrac{C_rP_r\ a^4}{J_2\  R_E^2\  \mu} \right)^{1/2} (1-e^2)^{5/4} \sqrt{e}\left(\dfrac{c_s^2\ \left(6+\sqrt{6+\dfrac{20}{3}\dfrac{a^2 \ n_S}{J_2\  R_E^2\ n}(1-e^2)^2} \right)}{\left\lvert6+20\dfrac{a^2 \ n_S}{J_2\  R_E^2\ n}(1-e^2)^2-4\sqrt{6+\dfrac{20}{3}\dfrac{a^2 \ n_S}{J_2\  R_E^2\ n}(1-e^2)^2}\right\rvert}\right)^{1/2}.
\end{align}

We now discuss the $j=-1,k=-1$ case. The location of the approximate equilibria in the $(\eta,\alpha)$-plane are the same, but because of the definition of $\alpha_j$ and of formula (\ref{cosrel}), the resonant curves in the $(i,e)$-plane are symmetric to the ones with $j=1$ with respect to the line $i=90^\circ$.

Moreover, the terms $C_{1,-1}$ and $C_{-1,-1}$ satisfy the following property:
\begin{equation}
C_{-1,-1}(\alpha_{-1})=\dfrac{\sin^2 \frac{i_s}{2}}{\cos^2 \frac{i_s}{2}}\ C_{1,-1}(\alpha_1\to\alpha_{-1}).    
\end{equation}
This implies that the amplitudes $\Delta_{-1,-1}$ verify
\begin{equation}
    \Delta_{-1,-1,\ell}(\alpha_{-1})=\dfrac{\sin \frac{i_s}{2}}{\cos \frac{i_s}{2}}\Delta_{1,-1,\ell}(\alpha_1\to\alpha_{-1}),
\end{equation}
so that the maximum resonance width for the case $j=-1,k=-1$ is five times smaller than the case $j=1,k=-1$. This phenomenon can be appreciated in Figures \ref{SRP1m1eccinc} and \ref{SRP1m1}, and in the numerical tests of the following section. We remark that the formulas for $\Delta_{1,-1,-1}$ are singular for $e^*$ and $i^*$ so that $\alpha_j(e^*,i^*)=\alpha_j(E_{-1})$. The numerical tests of Section \ref{FLIsec} show that these formulas are accurate outside a neighborhood of $e^*$ whose size depends on the area-to-mass ratio and the resonance under consideration.

\begin{figure}

\centering
 \includegraphics[width=1\textwidth]{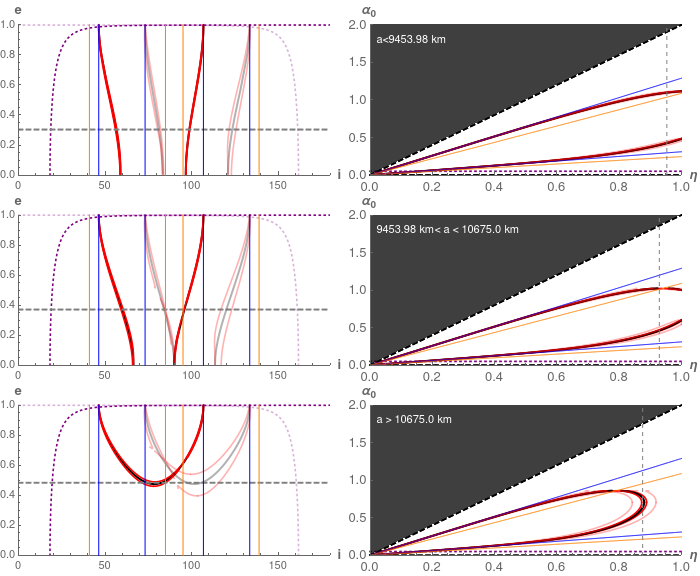} 
 \caption{Resonant curves for the resonance $j=\pm1, k=1$ in the $(i,e)$-plane (left column) and in the $(\hat{\eta},\alpha)$-plane (right column). The choice of the colors and the represented curves are the same as in Figures \ref{SRP1m1} and \ref{SRP1m1eccinc}. The plots depict the three different qualitative scenarios for this resonance: the top row corresponds to the case $z_p\leq(8\sqrt{10}+10)/45$, or $a\leq 9453.98$ km; the middle row to $(8\sqrt{10}+10)/45<z_p\leq 6/5$, or $9453.98$ km $<a\leq 10675$ km; the bottom row to $z_p> 6/5$, or $a>10675$ km.}
    \label{SRP11}
\end{figure}

Finally, we consider $j=\pm1,k=1$. In these two cases the upper branch admits a maximum point $E_1$, while there are no extremal points for the lower branch for all values of $z_p$. Moreover, the two branches can meet at a junction point $J$ provided that the semi-major axis is large enough. The key value associated to the appearance of $E_1$ is $z_p=10+8\sqrt{10}/45$, i.e. $a\simeq9454$ while the two branches of the resonant curves meet for $z_p>6/5$, i.e. for $a>10675.0$ km.
The three possible qualitative pictures under the variation of the semi-major axis are depicted in Figure (\ref{SRP11}).

Depending on the dynamical regime, there could be up to two PLE. For a fixed value of $\alpha_{\pm 1}$ we can have either two equilibria defined by the upper branch or one equilibrium point for each branch. From the study of the sign of $[C_{1,1}\Z'']_{\eta=\heta}$ one can conclude that stable points are $A_{1,N},B_{1_S}$ and $A_{-1}$.

The resonance width for the $j=1, k=1$ case is obtained as usual using formula (\ref{amp}) and substituting $\alpha_{ 1}$ with the expression for a branch of the resonant curve. One has that

\begin{align}
    \Delta_{1,1,1}=&\ \sqrt{2}\sqrt{\dfrac{A}{m}}\left(\dfrac{C_rP_r\ a^4}{J_2\  R_E^2\  \mu} \right)^{1/2} \eta^{5/2} (1-\eta^2)^{1/4}\left(\dfrac{s_s^2\ \left\lvert 6-\sqrt{6-5 z_p \eta^4}\right\rvert}{\lvert 6-15z_p\eta^4+4\sqrt{6-5z_p\eta^4}\rvert}\right)^{1/2}\\
    =&\  \sqrt{2}\sqrt{\dfrac{A}{m}}\left(\dfrac{C_rP_r\ a^4}{J_2\  R_E^2\  \mu} \right)^{1/2} (1-e^2)^{5/4} \sqrt{e}\left(\dfrac{s_s^2\ \left\lvert 6-\sqrt{6-\dfrac{20}{3}\dfrac{a^2 \ n_S}{J_2\  R_E^2\ n}(1-e^2)^2} \right\rvert}{\left\lvert6-20\dfrac{a^2 \ n_S}{J_2\  R_E^2\ n}(1-e^2)^2+4\sqrt{6-\dfrac{20}{3}\dfrac{a^2 \ n_S}{J_2\  R_E^2\ n}(1-e^2)^2}\right\rvert}\right)^{1/2} ,\\
     \Delta_{1,1,-1}=&\ \sqrt{2}\sqrt{\dfrac{A}{m}}\left(\dfrac{C_rP_r\ a^4}{J_2\  R_E^2\  \mu} \right)^{1/2} \eta^{5/2} (1-\eta^2)^{1/4}\left(\dfrac{s_s^2\ \left( 6+\sqrt{6-5 z_p \eta^4}\right)}{\lvert6-15z_p\eta^4-4\sqrt{6-5z_p\eta^4}\rvert}\right)^{1/2}\\
    =&\  \sqrt{2}\sqrt{\dfrac{A}{m}}\left(\dfrac{C_rP_r\ a^4}{J_2\  R_E^2\  \mu} \right)^{1/2} (1-e^2)^{5/4} \sqrt{e}\left(\dfrac{s_s^2\ \left(6+\sqrt{6-\dfrac{20}{3}\dfrac{a^2 \ n_S}{J_2\  R_E^2\ n}(1-e^2)^2} \right)}{\left\lvert6-20\dfrac{a^2 \ n_S}{J_2\  R_E^2\ n}(1-e^2)^2-4\sqrt{6-\dfrac{20}{3}\dfrac{a^2 \ n_S}{J_2\  R_E^2\ n}(1-e^2)^2}\right\rvert}\right)^{1/2} 
\end{align}
  
We conclude our analysis with the case $j=-1,k=1$. We can proceed as for $k=-1$ and we remark that
\begin{equation}
C_{-1,1}(\alpha_{-1})=\dfrac{\cos^2 \frac{i_s}{2}}{\sin^2 \frac{i_s}{2}}\ C_{1,-1}(\alpha_1\to\alpha_{-1}).    
\end{equation}
This implies that the amplitudes $\Delta_{-1,-1}$ verify
\begin{equation}
    \Delta_{-1,1,\ell}(\alpha_{-1})=\dfrac{\cos \frac{i_s}{2}}{\sin \frac{i_s}{2}}\Delta_{1,1,\ell}(\alpha_1\to\alpha_{-1}),
\end{equation}
so that the maximum resonance width for the case $j=-1,k=1$ is roughly five times the one for $j=1,k=1$.

In Appendix \ref{appC} we provide the approximate location of the six SRP resonances in the $(i,e)$-plane. To take into account all the different qualitative behaviours we use the key values defined previously in the section. The resulting plots are collected in Figure \ref{allie}, where we expressed the key values in terms of the semi-major axis of the orbit $a$, given in km. We notice that many resonant curves intersect at $i=90^{\circ}$ for $a>10133.2$. This region is investigated numerically in Section \ref{FLIsec}.
The shift between the approximated and real location of the equilibria depends on the value of the smallness parameter $\epsilon$, and, as a consequence, on $a$ and $\frac{A}{m}$. If $\epsilon$ is small enough the estimated resonances are very close to the real ones, with the latter slightly separating close to the $e=0$ line; if $\epsilon$ is not sufficiently small the curve representing the stable and unstable resonant point will diverge from the estimated one almost everywhere on the $(i,e)$ plane. This situation is portrayed in Figure 12 of \cite{alecol19}, which presents the location of the non-approximated resonant curves in the ($i,e$)-plane for $\frac{A}{m}=1$ and $20 $ m\textsuperscript{2}/kg at $a=10078$ km. In the left plot, which refers to $\frac{A}{m}=1$, $\epsilon \simeq 0.0026$ and one can see that the non-approximated resonant curves are very close to the one presented here. On the other hand, from the right plot, which referes to $\frac{A}{m}=20$, $\epsilon\simeq0.1$ and it is evident that the non-approximated resonant curves strongly diverge from the approximated ones. Some of the branches extend without intersecting the line $e=0$. In Section \ref{FLIsmall} we will study the behaviour close to $e=0$ using Fast Lyapunov Indicators and we will describe how the phase portrait qualitatively varies as a function of the area-to-mass ratio.

\subsection{Comparison of Resonances Strength}\label{resostr}
\begin{table}
\begin{center}
\begin{minipage}{\textwidth}
\begin{tabular}{@{}rrrllrrrrrrcc@{}}
\toprule
$j$ & $k$ & $\ell$ & \multicolumn{2}{c}{$\frac{A}{m}=10$ m\textsuperscript{2}/kg} & \multicolumn{2}{c}{$\frac{A}{m}=1$ m\textsuperscript{2}/kg} & $a$ &$e$ & $i$ (deg) & $a^*$ & Sing. & Period (yrs)\\
\midrule
& & &$\Delta_{\text{max}}$ &  $\delta q$ (km)  & $\Delta_{\text{max}}$ & $\delta q$ (km)  & & & & &(y/n)  & $\frac{A}{m}=1$ m\textsuperscript{2}/kg\\
\midrule
0 & -1 & 1 & 0.04869 & 1044.11 &0.01540 & 327.5 & 2.53 & 0.605  & 38.6 & 3.60 & n & 14.0405\\
0& -1 & -1 & 0.04869 &1044.11 &0.01540 & 327.5& 2.53 & 0.605 & 141.4 & 3.60 & n & 14.0405\\
0& 1& $\pm$ 1 & 0.06801&1688.07& 0.02151 &500.29 &1.80 & 0.445 &90.0 &1.80 &y & 19.7702\\
1& -1 &1 & 0.04940 &1058.51& 0.01562 &332.11 &2.59& 0.614& 128.3 &4.51 & n & 9.8886\\
1& -1 & -1&  0.17779 &4572.57& 0.05622 &1210.41 & 2.50 & 0.600 & 0.0 &2.50 & y & 7.1061\\
1& 1 & $\pm$ 1 & 0.02533 & 567.29&0.00801 &178.47& 1.96& 0.489 & 78.5&1.96 &y & 54.8171\\
-1& -1 & 1& 0.01025 & 217.76& 0.00324 & 68.84 &2.59 & 0.614 &51.7 &4.51 & n & 47.6678\\
-1& -1 & -1&0.03688 &788.31&0.01166 &248.09 & 2.50& 0.600 &180.0 & 2.50&y & 34.2551\\
-1& 1 & $\pm$ 1& 0.12210& 3470.23& 0.03861& 871.54& 1.96& 0.489 & 101.5& 1.96 & y & 11.3717\\

\bottomrule
\end{tabular}
\end{minipage}
\end{center}
\caption{Maximum amplitude of the resonant width $\Delta_{j,k}$, together with the location of the center of the resonance which realizes the maximum in terms of the classical Keplerian orbital elements. Note that $a$ and $a^*$ are provided in Earth's radii. The last column provides the fundamental period at the stable equilibrium for an object with area-to-mass equal to $1$ m\textsuperscript{2}/kg, expressed in years. See Section \ref{resostr} for more details. }\label{widtable}
\end{table}
Table \ref{widtable} provides the maximum amplitudes of the islands of stability for each of the six most relevant terms in the SRP expansion for two high values of the area-to-mass ratio, 1 and 10 m\textsuperscript{2}/kg, calculated by numerically maximizing the expression for $\Delta$ provided above. The range of $\eta$ over which we maximize $\Delta$ depends on the specific resonance under consideration, with the maximum value for $\eta$ depending on one of the aformentioned key values. We once again proceed as in \cite{breiter_aps} by including the location of the approximated center of the resonant island in terms of the usual orbital elements, and the semi-amplitude of the variation of the perigee, $\delta q$ (the total variation being of $2 \delta q$). Moreover, for each resonance we also include the value $a^*$ which corresponds to the maximum semi-major axis which allows an equilibrium point to be in the admissible region of the eccentricity, i.e. the eccentricity is greater than the critical value which corresponds to a perigee lower than the Earth's radius. We highlight the cases in which we encountered a singularity due to the presence of an extremal point, where $\Z''=0$ and the equations for $\Delta$ are singular, by a y/n flag in the second to last column of the table. In this situation we considered as $\Delta_{\text{max}}$ the value of $\Delta$ assumed at $a^*$. Finally, in the last column we provide the fundamental period at the stable equilibrium for the values of the orbital elements which realize the maximum value of $\Delta_{j,k}$ for an object with area-to-mass equal to $1$ m\textsuperscript{2}/kg, computed using Eqs. (\ref{funfreq}) and (\ref{funper}).

From Table \ref{widtable} it is straightforward to conclude that the strongest resonances are the $(1,-1)$ and the $(-1,1)$, followed by the $(0,1)$. The weakest resonance is the $(-1,-1)$. Nonetheless, if the area-to-mass is large, even the weakest resonance could cause a large change in the perigee distance. For example, the resonance $(-1,-1)$ induces a total variation in the perigee of $2\delta q\simeq 450$ km for a debris with $\frac{A}{m}=10 $ m\textsuperscript{2}/kg.
We remark that this "hierarchy" of SRP resonances is confirmed by the Fast Lyapunov Indicators maps in Appendix \ref{appC}. For example, the resonance $(-1,-1)$ is confirmed to be the weakest, and it is barely visible in the maps with $\frac{A}{m}=1 $ m\textsuperscript{2}/kg. We conclude by noting that the strongest resonances are also the fastest, acting on timescales of the order of a few years, while the weakest have a long-term effect, with a fundamental period of the order of decades.

\section{Numerical Validation}\label{FLIsec}
\begin{figure}
\centering
\begin{subfigure}{.48\textwidth}
    \centering
    \includegraphics[width=1\textwidth]{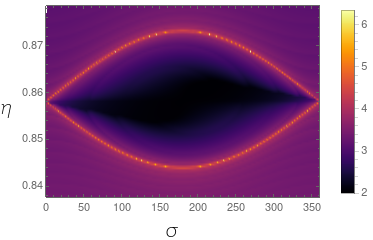}
\end{subfigure}
\begin{subfigure}{.48\textwidth}
    \centering
    \includegraphics[width=1\textwidth]{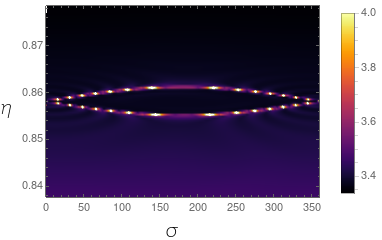}
\end{subfigure}
    \caption{\footnotesize Example of FLI plots depicting the phase portrait of the Hamiltonian $\mathcal{K}_{j,k}^*$ in terms of the resonant angle $\sigma$ and conjugated momentum $\eta$ for the $(1,-1)$ toy model (\textit{left}) and the $(-1,-1)$ toy model (\textit{right}), obtained using $a=2.064\ R_E$ , $\Omega=0$, $A/m=1$ m\textsuperscript{2}/kg and $\alpha_1=1.3$, which at $e=0$ corresponds to $i=113.578^\circ$ for $j=1$ and to $i=66.4218^\circ$ for $j=-1$. The FLI maps are obtained by integrating the non-autonomous 2 dof model for 30 years (\textit{left figure}) and 150 years (\textit{right figure}) with a timestep of 10 days.}
    \label{SRPm1_phase_space}
\end{figure}

The aim of this Section is to numerically validate the analytical approximations of the location of the equilibria, separatrix widths and periods of the previously described toy models. 
We compare the solutions obtained by using the toy models with the ones obtained by numerically integrating, using an Adam-Bashfort-Moulton scheme, the Cartesian equation of motion of a \textit{full model}, which includes all the main relevant perturbations to the two-body problem, i.e. higher-degree terms of the geopotential, the lunisolar perturbations and the SRP effect. By doing so we are actually comparing two different kinds of solutions, one expressed in \textit{mean elements}, and the other expressed in the classical Keplerian elements, including the fast variable $M$. In order to validate the analytical prediction, we require the Cartesian solution to differ only by the \textit{short periodic perturbations} which are expected to be present in a non-averaged model.
\color{black}
Moreover, we are going to make use of \textit{Fast Lyapunov Indicators}, or FLIs, \cite{flis}. FLIs are powerful chaos indicators which let one distinguish between chaotic and regular motions. In Appendix \ref{appB} we provide the formal definition of the FLIs for a generic dynamical system, not necessarily expressed in the Hamiltonian framework.
In the plots of this section darker colors refer to a stable motion, while brighter colors refer to unstable or chaotic dynamics. Bright curves will be associated to unstable equilibria and \textit{separatrices}, but can also be used to highlight the effect of a singularity. We will focus on this aspect later in this section, when we describe the behaviour of FLI maps close to $e=0$. Therefore, we will use FLIs to draw phase portraits and bifurcation diagrams. In order to compute the FLI maps we integrate the Hamiltonian equations of motion and the variational equation associated to $\H_{j,k}$.

We use FLIs to validate the bifurcation plots presented in the previous Section and, in particular, we analyse some of the limit cases that were neglected during our previous study, such as the case in which $\alpha_j$ is close to an extremal of a resonant curve or the case of a large area-to-mass ratio value which results in a not small value of the parameter $\epsilon$. We use FLIs to describe the dynamics in the neighborhood of the singularity $e=0$. Finally, we study numerically how some of the SRP resonances overlap in the case of polar orbits. 

\subsection{Influence of the area-to-mass ratio parameter}

\begin{figure}
\centering
\begin{subfigure}{.48\textwidth}
    \centering
    \includegraphics[height=0.9\textwidth]{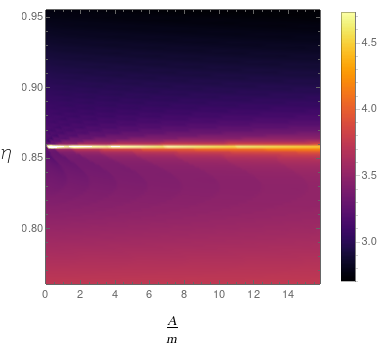}
\end{subfigure}
\begin{subfigure}{.48\textwidth}
    \centering
    \includegraphics[height=0.9\textwidth]{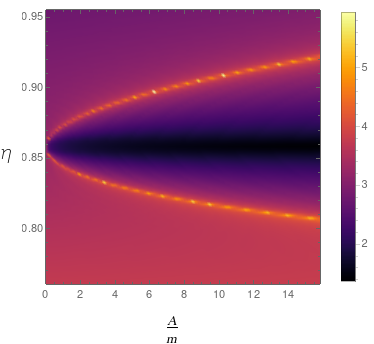}
 \end{subfigure}

\begin{subfigure}{.48\textwidth}
    \centering
    \includegraphics[height=0.9\textwidth]{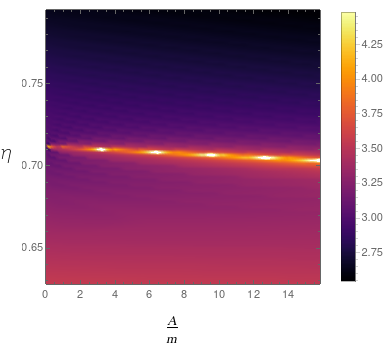}
\end{subfigure}
\begin{subfigure}{.48\textwidth}
    \centering
    \includegraphics[height=0.9\textwidth]{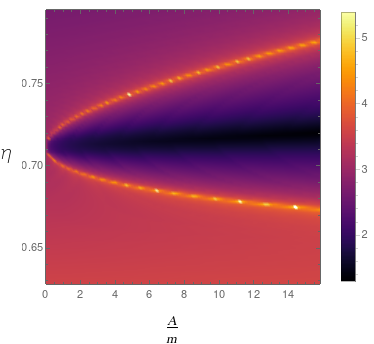}
\end{subfigure}
\caption{\footnotesize Resonance $(1,-1)$. Location of the unstable equilibrium point (left column) and resonance width of the stable equilibrium point (right column), for $\alpha_1=1.4$, $\Omega=0$ and $\omega$ so that $\sigma_{1,-1}=0$ (left) and $\sigma_{1,-1}=180^\circ$ (right), using FLIs. The top plots correspond to $a=2.064 \ R_E$, where $\epsilon=0.1$ for $A/m=12.82$ m\textsuperscript{2}/kg. The bottom plots correspond to $a=3.5\ R_E$, where $\epsilon=0.1$ for $A/m=1.55$ m\textsuperscript{2}/kg.}
\label{locationwidth}
\end{figure}

We start by drawing FLI maps describing the SRP resonances for various values of the area-to-mass ratio parameter. We recall that the approximations from Section \ref{model} are only valid for values of $a$ and $A/m$ so that the dimensionless quantity $$\epsilon=\frac{C_r P_r a}{J_2 R_E^2 n^2}\frac{A}{m}$$ is small, let us say $\epsilon<0.1$ for reference. We can confirm this by solving numerically the problem defined by the canonical equations associated to the non-autonomous 2 degrees-of-freedom Hamiltonian toy-models $\H_{j,k}$ from Eq. (\ref{toymodels}) and by computing the FLIs to draw phase portraits and bifurcation diagrams. Using this procedure one can appreciate the deviations from the predicted values without the need of computing explicit formulas to draw complex bifurcation plots.

Figure \ref{SRPm1_phase_space} shows an example of phase portraits obtained using FLI maps generated by integrating the canonical equations corresponding to $\H_{j,k}$ in Eq. (\ref{toymodels}), for the resonances $(1,-1)$ and $(-1,-1)$, at $a=2.064\ R_E$ and for $A/m=1$ m\textsuperscript{2}/kg. In this situation one has that $\epsilon=7.8\times 10^{-3}$, and the smallness condition is satisfied. We remark that, as a consequence, the formulas of the previous Section yield a good approximation for both the location of the equilibrium points and the resonance width. Moreover, Figure \ref{SRPm1_phase_space} confirms the results from the previous section, according to which the separatrix width of the resonance $(-1,-1)$ is about five times smaller than the one of the resonance $(1,-1)$.

The FLI maps in Figure \ref{locationwidth} show how the location of the unstable equilibrium point and the resonance width vary with respect to the area-to-mass ratio parameter for the resonance $(1,-1)$. The dark region in the plots on the right column corresponds to initial conditions taken sufficiently close to the stable equilibrium point. The panels of Figure \ref{locationwidth} are obtained from the FLIs for $\sigma_{1,-1}=0$ or $\sigma_{1,-1}=180^\circ$ and by varying the area-to-mass ratio. The  plots in the first row correspond to an object which has the same initial conditions as in the left panel of Figure \ref{SRPm1_phase_space}. Since for $a=2.064\ R_E$ the parameter $\epsilon$ is small for almost all values of $A/m\in[0,16]$ m\textsuperscript{2}/kg, we have that the location of both the unstable and stable equilibrium points does not change with the variation of $A/m$ while the resonant width grows as the square root of the area-to-mass ratio, as mentioned in Section \ref{model}. On the other hand, if the smallness condition for $\epsilon$ is not satisfied we found a small deviation from the approximated location of the equilibria, which grows linearly with the area-to-mass ratio, such as in the bottom panels of Figure \ref{locationwidth}, where $a=3.5\ R_E$ and $\epsilon=0.1$ when $A/m\simeq 1.55$ m\textsuperscript{2}/kg.

We considered the case of one isolated equilibrium point to avoid any \textit{merging} with a nearby resonant island which could have affected our validation of the location and amplitude formulas. The case of the merging of nearby resonances associated to the same resonant term is treated using FLIs in the Section \ref{merging}.


\subsection{Comparison with the Cartesian solution}

\begin{figure}
    \centering
\includegraphics[width=\textwidth]{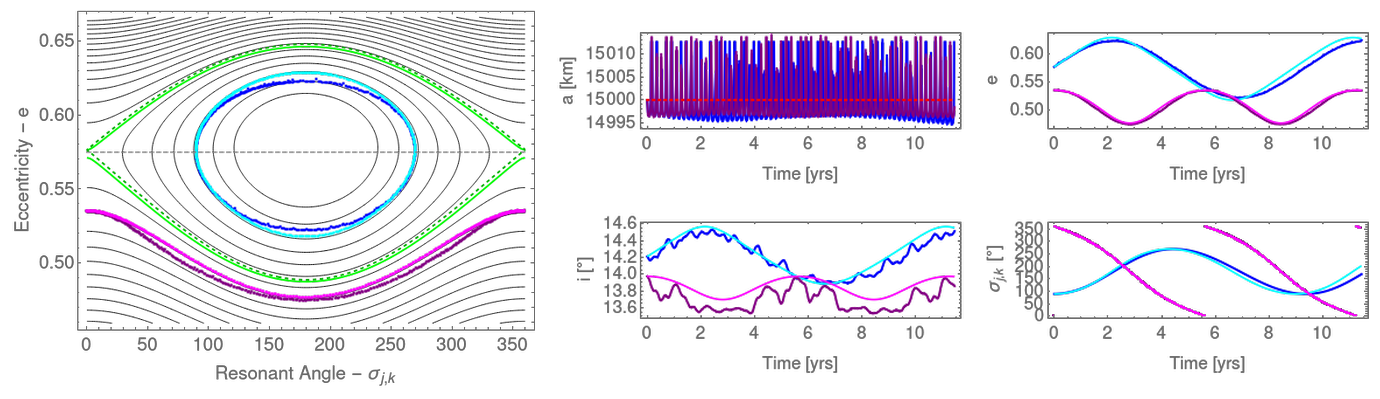}

    \caption{Resonance $(1,-1)$, $A/m=1$ m\textsuperscript{2}/kg, $a=15000$ km, $\alpha_1=0.025$, which correspond to $i=14.2089^\circ$ at the equilibrium $e_c=0.57637$. The initial conditions are such that $\sigma_{1,-1}(t_0)=90^\circ$, $e(t_0)=e_c$ (\textit{cyan and blue curves}), and $\sigma_{1,-1}(t_0)=0^\circ$, $e(t_0)=0.534966$ (\textit{magenta and purple curves}). The cyan and magenta curves correspond to the Hamiltonian averaged solutions, while the blue and purple one correspond to the Cartesian solutions. It is evident that the Hamiltonian solutions yield satisfactory results in approximating the Cartesian ones, especially inside the resonant island.}
    \label{comp1}
\end{figure}

\begin{figure}
    \centering

 \includegraphics[width=\textwidth]{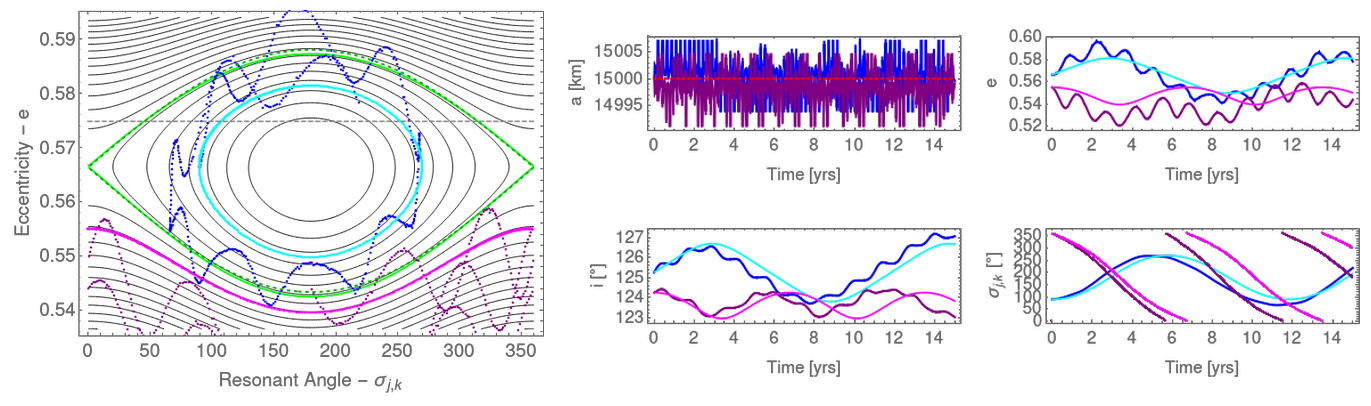}

    \caption{Resonance $(1,-1)$, $A/m=1$ m\textsuperscript{2}/kg, $a=15 000$ km, $\alpha_1=1.3$, which correspond to $i=125.261^\circ$ at the equilibrium $e_c=0.566314$. The initial conditions are such that $\sigma_{1,-1}(t_0)=90^\circ$, $e(t_0)=e_c$ (\textit{cyan and blue curves}), and $\sigma_{1,-1}(t_0)=0^\circ$, $e(t_0)=0.554996$ (\textit{magenta and purple curves}). Contrary to the case of Figure \ref{comp1}, the Hamiltonian solution does not provide the same level of accuracy in approximating the full solution. In particular, the circulating solution differ greatly in all the relevant components. The librating solutions exhibit large variation, but the error is not as large as in the circulating case.}
    \label{comp2}
\end{figure}

\begin{figure}
    \centering
 \includegraphics[width=\textwidth]{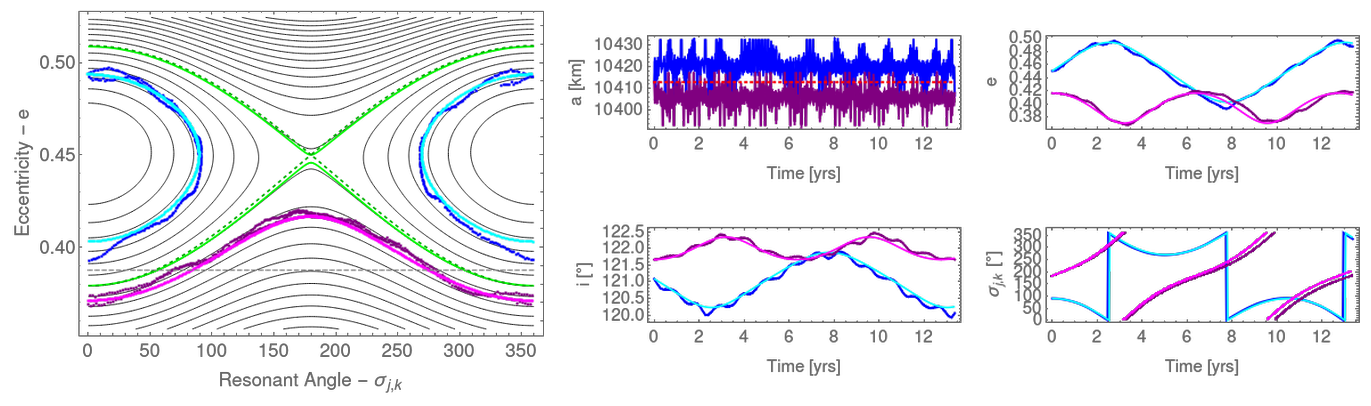}

    \caption{Resonance $(-1,1)$, $A/m=1$ m\textsuperscript{2}/kg, $a=10412.9$ km, $\alpha_{-1}=0.431911$, which correspond to $i=121.088^\circ$ at the equilibrium $e_c=0.45$. The initial conditions are such that $\sigma_{-1,1}(t_0)=90^\circ$, $e(t_0)=e_c$ (\textit{cyan and blue curves}), and $\sigma_{-1,1}(t_0)=180^\circ$, $e(t_0)=0.416316$ (\textit{magenta and purple curves}).}
    \label{comp3}
\end{figure}

\begin{figure}
    \centering

 \includegraphics[width=\textwidth]{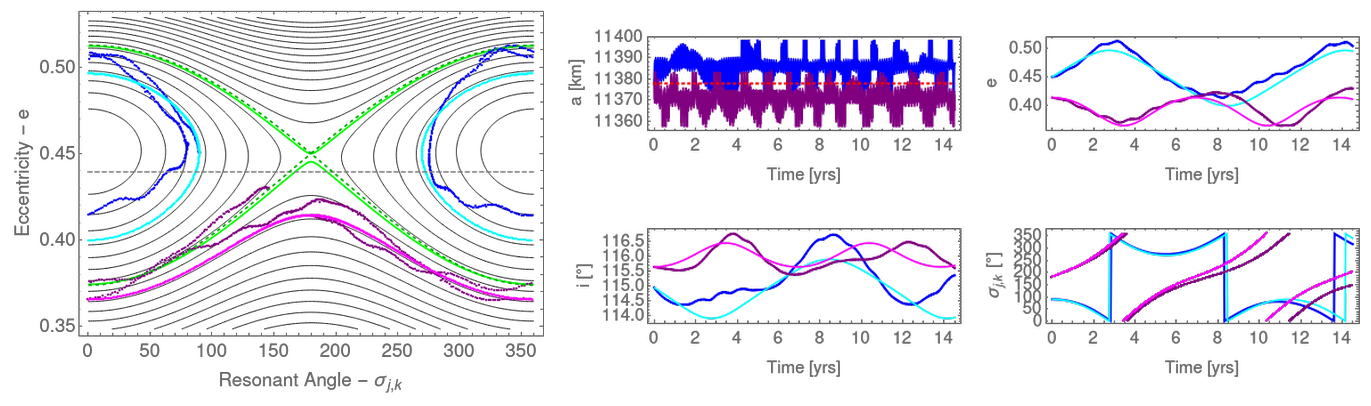}

    \caption{Resonance $(-1,1)$, $A/m=1$ m\textsuperscript{2}/kg, $a=11377.8$ km, $\alpha_{-1}=0.516343$, which correspond to $i=114.949^\circ$ at the equilibrium $e_c=0.45$. The initial conditions are such that $\sigma_{-1,1}(t_0)=90^\circ$, $e(t_0)=e_c$ (\textit{cyan and blue curves}), and $\sigma_{-1,1}(t_0)=180^\circ$, $e(t_0)=0.414003$ (\textit{magenta and purple curves}).}
    \label{comp3bis}
\end{figure}

Let us now compare the analytical approximation, obtained using a toy model, with the Cartesian solution, obtained by integrating the full Cartesian equations of motion, which are presented in Appendix \ref{appA}. We note that proceeding in this way is equivalent to comparing two different sets of elements, and we will complete any sets of initial conditions for the Hamiltonian model by including the mean anomaly $M(t_0)$ which will always be set to $0^\circ$. We choose the initial conditions for the angles such that $\Omega(t_0)=0^\circ$ and $\omega(t_0):=\sigma_{j,k}(t_0)-kM_S(t_0)$. 

\begin{figure}
    \centering

 \includegraphics[width=\textwidth]{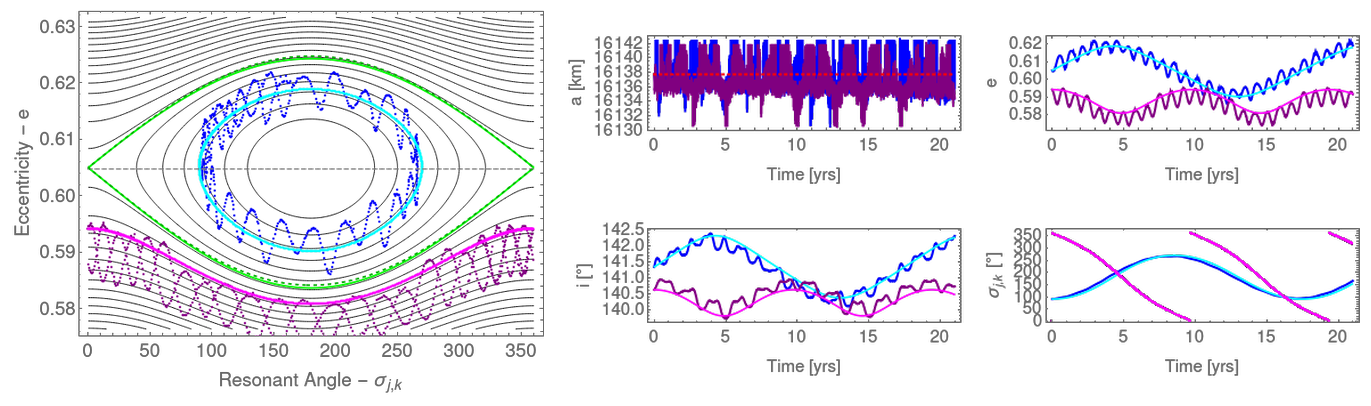}

 \caption{Resonance $(0,-1)$, $A/m=1$ m\textsuperscript{2}/kg, $a=16136.7$ km, $\alpha_0=-0.621694$, which correspond to $i=38.666^\circ$ at the equilibrium $e_c=0.605$. The initial conditions are such that $\sigma_{0,-1}(t_0)=90^\circ$, $e(t_0)=e_c$ (\textit{cyan and blue curves}), and $\sigma_{0,-1}(t_0)=0^\circ$, $e(t_0)=0.594131$ (\textit{magenta and purple curves}). The Hamiltonian approximation does not differ much from the Cartesian librating solution, but there are some discrepancies in the case of the circulating one.}
    \label{comp4}
\end{figure}

\begin{figure}
    \centering

 \includegraphics[width=\textwidth]{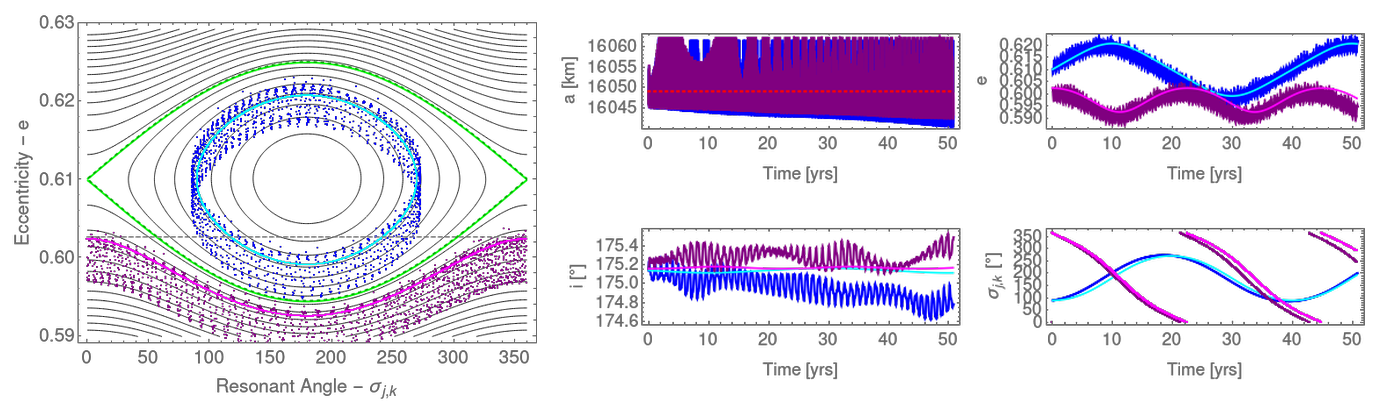}

    \caption{Resonance $(-1,-1)$, $A/m=1$ m\textsuperscript{2}/kg, $a=16049.1$ km, $\alpha_{-1}=0.00285$, which correspond to $i=175.139^\circ$ at the equilibrium $e_c=0.61$. The initial conditions are such that $\sigma_{-1,-1}(t_0)=90^\circ$, $e(t_0)=e_c$ (\textit{cyan and blue curves}), and $\sigma_{-1,-1}(t_0)=0^\circ$, $e(t_0)=0.602301$ (\textit{magenta and purple curves}).}
    \label{comp5}
\end{figure}

\begin{figure}
    \centering

 \includegraphics[width=\textwidth]{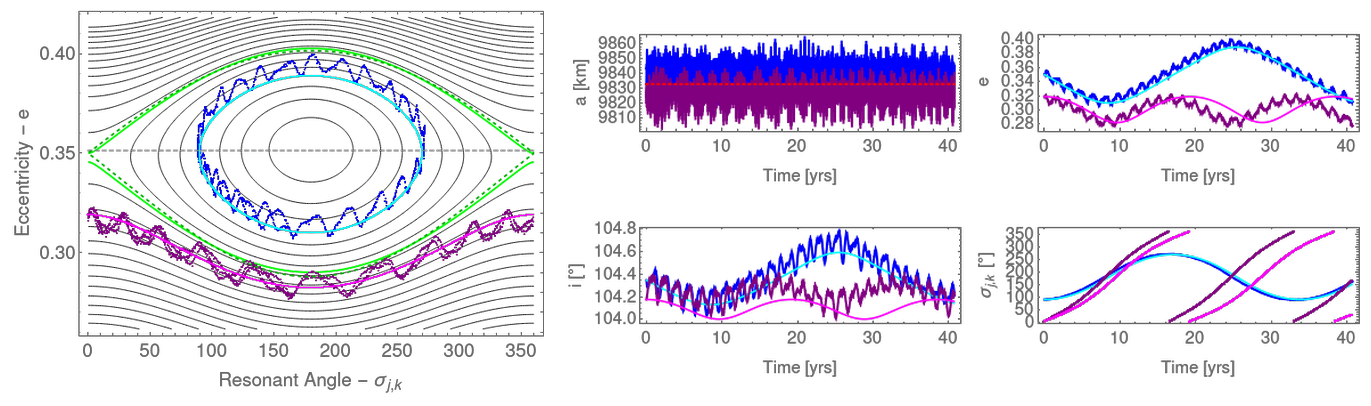}

    \caption{Resonance $(0,1)$, $A/m=1$ m\textsuperscript{2}/kg, $a=9832.69$ km, $\alpha_0=0.232115$, which correspond to $i=104.347^\circ$ at the equilibrium $e_c=0.35$. The initial conditions are such that $\sigma_{0,1}(t_0)=90^\circ$, $e(t_0)=e_c$ (\textit{cyan and blue curves}), and $\sigma_{0,1}(t_0)=0^\circ$, $e(t_0)=0.318829$ (\textit{magenta and purple curves}).}
    \label{comp6}
\end{figure}

\begin{figure}
    \centering

 \includegraphics[width=\textwidth]{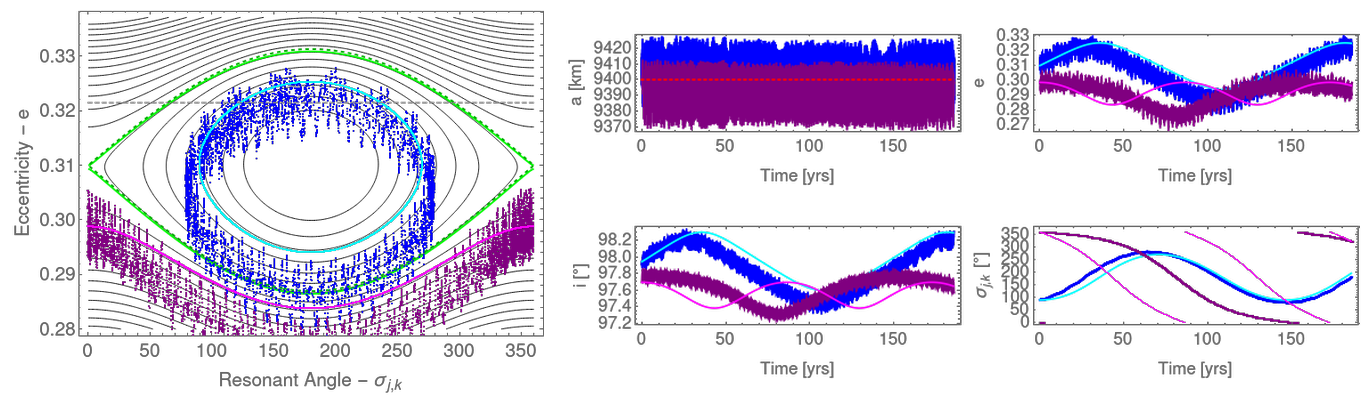}

    \caption{Resonance $(1,1)$, $A/m=1$ m\textsuperscript{2}/kg, $a=9400$ km, $\alpha_1=1.08211$, which correspond to $i=97.9426^\circ$ at the equilibrium $e_c=0.31$. The initial conditions are such that $\sigma_{1,1}(t_0)=90^\circ$, $e(t_0)=e_c$ (\textit{cyan and blue curves}), and $\sigma_{1,1}(t_0)=0^\circ$, $e(t_0)=0.298719$ (\textit{magenta and purple curves}).}
    \label{comp7}
\end{figure}

The numerical comparisons in Figures \ref{comp1} to \ref{comp7} include the solutions represented on the $(\sigma,e)$ plane in the neighborhood of the PLE, together with the time series of the semi-major axis, eccentricity, inclination and resonant angle. In the left panels, we plot also the theoretical separatrices estimated from the pendulum model (\textit{dark green dashed curves}), the separatrices computed numerically from the toy model (\textit{light green curves}), and the critical eccentricity (\textit{dashed grey line}).

The first two comparisons, presented in Figures \ref{comp1} and \ref{comp2}, are focused on the resonance $(1,-1)$ and they correspond to the two dynamical regimes highlighted in Figure \ref{exampleres}, where the mean semi-major axis is $a=15000$ km. In particular, Figure \ref{comp1} illustrates the phase portrait around the stable equilibrium point arising by fixing $\alpha_1=0.025$, which corresponds to an inclination of $i\simeq 14.21^\circ$ at the equilibrium point $e=0.57637$. Figure \ref{comp1} shows that the Hamiltonian solution provides a good approximation of the full model, in particular in the case of the circulating solution, inside the island of stability. 

The results are very different in the case of $\alpha_1=1.3$ (see Figure \ref{comp2}). This choice of the dynamical regime corresponds to an inclination of $i=125.261^\circ$ at the equilibrium $e=0.566314$. Since the inclination is large, the $(1,-1)$ SRP term is not dominant as it was in the previous case (see Figure \ref{magni}), making the short periodic perturbations due to the other SRP resonant terms more prominent. Nonetheless, in the librating case the averaged solution is still able to approximate the Cartesian dynamics, while in the case of the circulating orbit the Cartesian solution lags considerably behind the approximate solution, as it is evident from the plots on the right side of Figure \ref{comp2}.

\begin{figure}
    \centering

 \includegraphics[width=\textwidth]{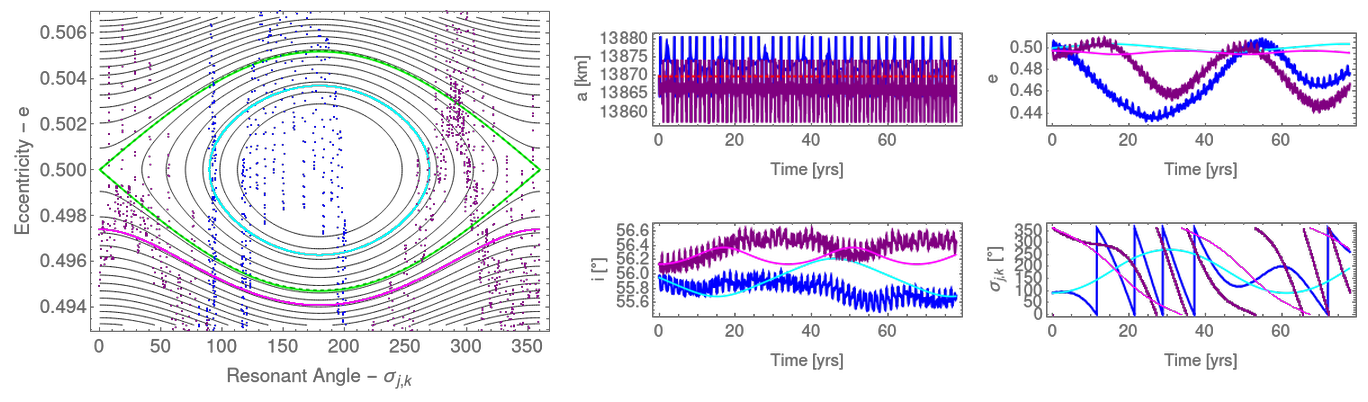}

    \caption{Resonance $(-1,-1)$, $A/m=1$ m\textsuperscript{2}/kg, $a=13869.7$ km, $\alpha_{-1}=1.35094$, which correspond to $i=55.95^\circ$ at the equilibrium $e_c=0.5$. The initial conditions are such that $\sigma_{-1,-1}(t_0)=90^\circ$, $e(t_0)=e_c$ (\textit{cyan and blue curves}), and $\sigma_{-1,-1}(t_0)=0^\circ$, $e(t_0)=0.497373$ (\textit{magenta and purple curves}).}
    \label{comp8}
\end{figure}

Another example is given by Figure \ref{comp3}, dedicated to the second most relevant resonance, i.e. the one associated to the $(-1,1)$ SRP term. In this case the Hamiltonian solutions provide a good approximation for both circulating and librating solutions. If one increases the semi-major axis and slightly changes the initial conditions, as in Figure \ref{comp3bis}, the analytical predictions lose accuracy in approximating the Cartesian solutions. However, the dynamics revealed by the two models is topologically equivalent.

Figures \ref{comp4} through \ref{comp7} are focused on the weaker resonances, which all exhibit similar features: the Hamiltonian approximation is satisfactory in approximating the $e,i$ and $\sigma_{j,k}$ time series, at least for librating orbits. The numerical simulations confirm that these resonances act on large timescales, with variations whose period is of the order of several decades. 

We conclude this subsection by showing an example, depicted in Figure \ref{comp8}, where the analytical approximation does not hold. The resonance under study is the $(-1,-1)$, and the initial conditions are taken close to the secular lunisolar resonance $2\dot{\omega}+\dot{\Omega}=0$. The plot of the time series of the eccentricity, inclination and resonant angle clearly show that in this situation the Hamiltonian approximation strongly diverges from the Cartesian solution. We recall that each lunisolar resonance corresponds to one of the critical inclinations, where the theory from Section \ref{model} is not applicable. In Section \ref{merging} we will numerically describe the behaviour of the solutions close to the neighborhood of a region where $\mathcal{Z}^{''}=0$, i.e. close to the degenerate lines which separate the resonant curves in two branches of opposite stability.

We close this subsection with the following remark. A source of strong chaos in the SRP dynamics can be found in the regions where two distinct resonant curves overlap, as one can deduce from the FLI maps of Appendix \ref{appC}. In Section \ref{overlappi} we will focus on the overlapping of SRP resonances for polar orbits, i.e. orbits with $i=90^\circ$.

\subsection{Merging of nearby resonant islands}\label{merging}

\begin{figure}
    \centering
    \begin{subfigure}{0.45\textwidth}
    \includegraphics[width=\textwidth]{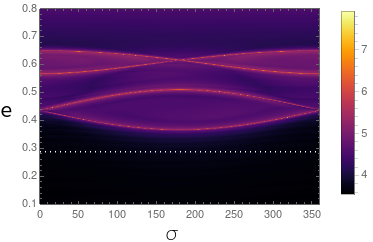}
    \end{subfigure}
    \begin{subfigure}{0.45\textwidth}
    \includegraphics[width=\textwidth]{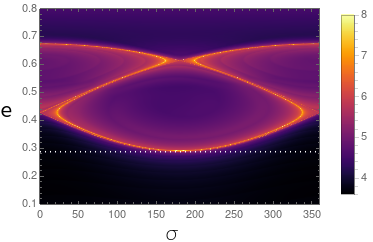}
    \end{subfigure}
    
    \begin{subfigure}{0.45\textwidth}
    \includegraphics[width=\textwidth]{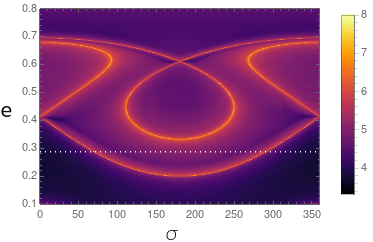}
    \end{subfigure}
    \begin{subfigure}{0.45\textwidth}
    \includegraphics[width=\textwidth]{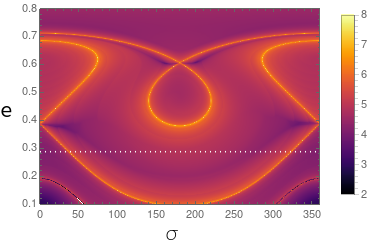}
    \end{subfigure}
    
    \begin{subfigure}{0.45\textwidth}
    \includegraphics[width=\textwidth]{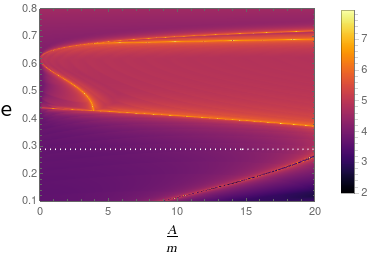}
    \end{subfigure}
    \begin{subfigure}{0.45\textwidth}
    \includegraphics[width=\textwidth]{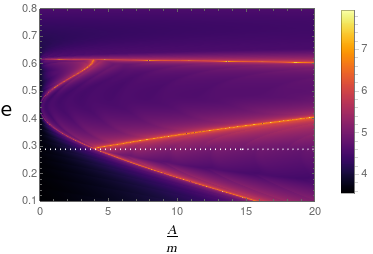}
    \end{subfigure}
    \caption{This figure depicts the phenomenon of the merging of resonances under the variation of the area-to-mass ratio parameter, using FLIs. The test case is the resonance $(0,1)$, with $a=1.405\ R_E$, $\Om=0^\circ$ and $\alpha_0=0.305$, which corresponds to $i=72.24^\circ$ at $e=0$. Propagation time: 75 years, timestep: 10 days. The first four plots show the phase portraits in the $(\sigma_{0,1},e)$ plane, for $A/m=1,4,9$ and $16$ m\textsuperscript{2}/kg, while, in the last row, bifurcations plots show how the equilibria and sepratrices behave under the variation of the area-to-mass ratio, up to $20$ m\textsuperscript{2}/kg. These two plots show how the sections obtained intersecting the phase portraits with the lines $\sigma_{0,1}=0^\circ$ (\textit{left}), or $\sigma_{0,1}=180^\circ$ (\textit{right)}, vary for different values of the $A/m$ parameter. The white dotted line corresponds to the critical value of the eccentricity, which in this case is equal to $e_{crit}\simeq 0.288$.}
    \label{merge1}
\end{figure}

Let us consider a resonance whose resonant curve allows for an extremal point\footnote{This could happen for every resonant toy models except for the $(0,-1)$, which never allows for neither a maximum nor a minimum point.}, and without loss of generality let it be a maximum point. Let us fix a value of the integral $\alpha$ slightly smaller than the one corresponding to the maximum, a situation that we disregarded in the previous section due to the singularity at $\Z''=0$. In this situation one has two equilibrium points $\eta_{S}<\eta_{N}$, together with the corresponding amplitudes $\Delta_{S}$ and $\Delta_{N}$. We say that this two nearby resonances are \textit{separated} if $\text{Max}(\Delta_{S},\Delta_{N})<\eta_{N}-\eta_{S}$, while we say that they \textit{merge} otherwise. In the following we will show how two initially separated resonances could merge under the variation of  the area-to-mass ratio.

In the first example we are going to focus on the (0,1) resonance, which is the third strongest resonance, see Table \ref{widtable}. We fix the parameter $a=1.784\ R_E$ and $\alpha_0=0.245$, corresponding to an inclination of $i=75.82^\circ$ at $e=0$. In this situation $\epsilon<0.1$ if $A/m<22.97$ m\textsuperscript{2}/kg, allowing us to consider very large values of the area-to-mass ratio in the subsequent bifurcation maps without affecting, in principle, the location of equilibria. Figure \ref{merge1} depicts the corresponding phase portraits for four values of the area-to-mass ratio, namely $A/m=1,5,9,20$ m\textsuperscript{2}/kg, together with two bifurcation diagrams depicting the evolution of the $\sigma=0^\circ$ and $\sigma=180^\circ$ sections of the phase space under the variation of the area-to-mass ratio parameter, allowing us to describe how the location of the equilibria and the newly formed branches of the separatrices vary. In the $A/m=1$ m\textsuperscript{2}/kg case one has that 
\begin{align*}
    \eta_N= 0.898433\longrightarrow e_S=0.439111,\\
    \eta_S= 0.787744\longrightarrow e_N=0.616002,
\end{align*}
so that $\eta_N-\eta_S=0.111372$, or $e_N-e_S=0.176891$, and
\begin{align*}
    \Delta_N=0.0571462,\\
    \Delta_S=0.0574857.
\end{align*}
Therefore, the two resonances are indeed separated as the first FLI map of Figure \ref{merge1} shows. We remark that the formulas for $\Delta$ are accurate only if it is possible to approximate the dynamics using a pendulum model. This is true when the two resonances are separated, however the amplitude formulas fail to predict the actual resonance width when the two resonance islands get closer and closer to each other. We proceed with a qualitative description of the phase portraits under the variation of the area-to-mass ratio. First the two separatrices meet, resulting in a \textit{saddle connection} (\cite{arnold2013dynamical}, second panel of the top row of Figure \ref{merge1})\footnote{The exact value corresponding to the saddle connection is $A/m\simeq 3.7$ and it has been computed numerically.}. 
By increasing the value of the area-to-mass ratio the two islands of stability start to merge, resulting in a phase portrait which is radically different from the one of the simple pendulum. In particular, we notice that the two stable equilibrium points are surrounded by two separatrices which bound the evolution of the resonant angle. The separatrix enclosing the stable equilibrium at $\sigma=180^\circ$ quickly gets smaller as the area-to-mass ratio increases, with the equilibrium point getting closer to the unstable equilibrium at $\sigma=180^\circ$. The separatrix enclosing the equilibrium at $\sigma=0^\circ$ persists for higher values of the area-to-mass ratio without shrinking down. 

The second plot of the central row, which represents the phase portrait for $A/m=16 $ m\textsuperscript{2}/kg, shows a darker region surrounding the point $(\sigma,e)=(0,e^*)$, with $e^*\simeq 0$. This is due to the presence of the stable equilibrium point $e^*$, which was disregarded as a consequence of neglecting the SRP term in Eq. (\ref{soleta}), see \cite{alecol19}. 
The curve made of bright halos surrounding dark spots which surrounds it is not a separatrix and it will be addressed later in this Section.

As mentioned before, one cannot describe the phenomenon of the merging of resonances using a pendulum model such as the one from Eq. (\ref{apppend}), since $\mathcal{Z}^{''}$ is close to zero in the neighborhood of a degenerate line. Numerical tests show that a third degree model is not enough to describe the merging phenomenon for SRP semi-secular resonances, therefore we postpone the formal study of the merging of nearby resonances to a future work.

\begin{figure}
    \centering
    \begin{subfigure}{0.45\textwidth}
    \includegraphics[width=1\textwidth]{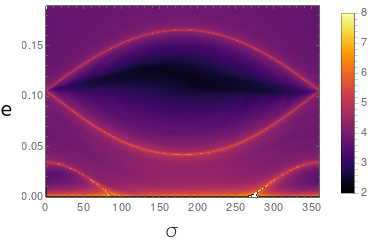}
    \end{subfigure}
        \centering
    \begin{subfigure}{0.45\textwidth}
    \includegraphics[width=1\textwidth]{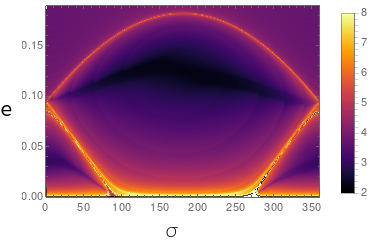}
    \end{subfigure}
    \begin{subfigure}{0.45\textwidth}
    \includegraphics[width=1\textwidth]{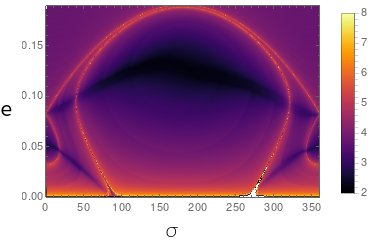}
    \end{subfigure}
    \begin{subfigure}{0.45\textwidth}
    \includegraphics[width=1\textwidth]{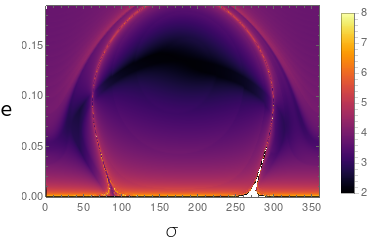}
    \end{subfigure}
    \begin{subfigure}{0.45\textwidth}
    \includegraphics[width=1\textwidth]{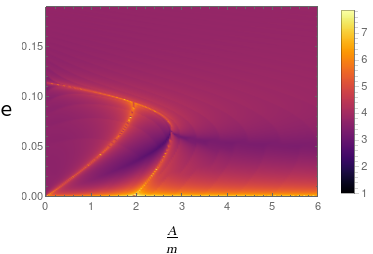}
    \end{subfigure}
    \begin{subfigure}{0.45\textwidth}
    \includegraphics[width=1\textwidth]{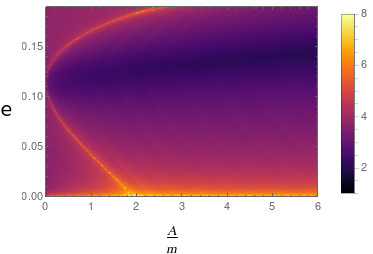}
    \end{subfigure}
    \caption{Resonance $(1,-1)$. Phase portrait in the neighborhood of $e=0$ using FLIs. The initial conditions are given by $a=1.589 R_E$, $\alpha_1=1.45$, $\Om=0$. In the first four plots (\textit{top two rows}) the area-to-mass corresponds respectively, to $1,1.9,2.5$ and $4$ m\textsuperscript{2}/kg. The bright lines emanating from $(e,\sigma)=(0,\pm 90^\circ)$ correspond to the $e=0$ level set and it highlights the singularity of the cylindrical action-angle variables. The last two FLI maps can be seen as bifurcation plots which show how the equilibrium points and the separatrices change under the variation of the area-to-mass ratio. }
    \label{smallecc}
\end{figure}
\subsection{Equilibrium points very close to $e=0$} \label{FLIsmall}

In this section we will briefly focus on equilibrium points which are estimated to be very close to the $e=0$ or, equivalently, $\eta=1$. As we mentioned before, using the current set of action-angle cylindrical coordinates, the problem presents virtual singularities. If one considers a value of $\alpha_j$ slightly smaller than the one corresponding to the intersection of the resonant curve with the line $\eta=1$, the corresponding pair of equilibrium points will be very close to $e=0$ line. The branch of the separatrix which is closest to that line might merge with the line itself, if the area-to-mass is big enough (see Figure \ref{smallecc}). We remark that in this situation the approximations are not as precise as in the previous cases, therefore the real locations of the equilibrium points are slightly shifted and the island of stability is not symmetric to the $\eta=\eta_u$ line, where $\eta_u$ is the value of $\eta$ which corresponds to an unstable solution. The actual location of the equilibria can be computed numerically by solving Eq. (\ref{soleta}). Figure 12 of \cite{alecol19} presents the location of the equilibria in the ($i,e$)-plane for two values of $A/m$, and it shows that the bigger the area-to-mass ratio, the bigger is the separation between the real locations of the stable and unstable equilibrium points and those estimated in the previous Section.
Figure \ref{smallecc} depicts how the phase portraits of an object subjected to the ($1,-1$) SRP resonance varies with the area-to-mass ratio, including two bifurcations diagrams in the bottom row. In the first map of Figure \ref{smallecc} one can observe a purple region at $\sigma=0$, very close to $e=0$, surrounded by a bright curve. The bright curve also appears in the subsequent plots, surrounding the stable equilibrium point. Such curve does not represent a branch of a separatrix, but it marks the level set of the toy model Hamiltonian $\mathcal{K}(\heta,\sigma)=\mathcal{K}(1,90^\circ)$, which in Keplerian elements correspond to the level set passing through $e=0$\footnote{We recall that $\eta=1$ is a singular point for the canonical equations, but it is not singular for the toy model Hamiltonians.}. The reason why this level set is detected as chaotic/unstable by the FLIs is because these indicators are sensitive to the evolution of the tangent vector of the system as a solution of the variational equation. Points which will tend to $e=0$ during their motion will experience a great variation in the tangent vector because of the singularity, resulting in extremely large values of the FLI, which might even cause an overflow, represented by the white spots near the points $(\sigma_{1,-1},e)=(\pm90^\circ,0)$ in the top four panels of Figure \ref{smallecc}. The dark purple region at $\sigma=0$ actually corresponds to a stable equilibrium point which is disregarded in the approximation given by Eq. (\ref{approxeta}) and which disappears, together with the unstable equilibrium point at $\sigma=0$, if one increases the area-to-mass ratio. The fourth plot in Figure \ref{smallecc} shows that for an area-to-mass ratio of 4 m\textsuperscript{2}/kg one can find a single stable equilibrium point surrounded by very wide orbits with bounded $\sigma$. 

\begin{figure}
    \centering
    \begin{subfigure}{0.45\textwidth}
    \includegraphics[width=1\textwidth]{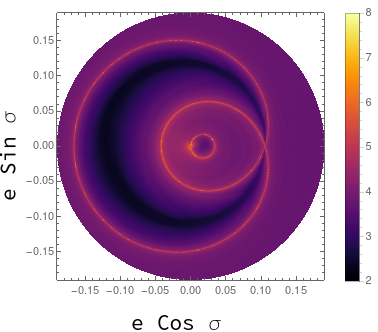}
    \end{subfigure}
        \centering
    \begin{subfigure}{0.45\textwidth}
    \includegraphics[width=1\textwidth]{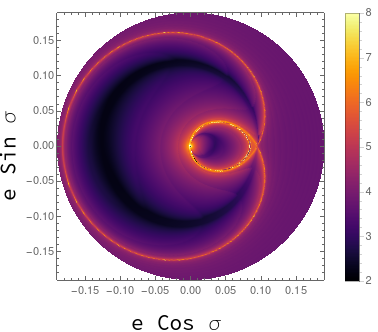}
    \end{subfigure}
    \begin{subfigure}{0.45\textwidth}
    \includegraphics[width=1\textwidth]{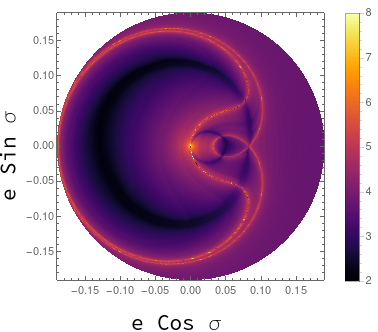}
    \end{subfigure}
    \begin{subfigure}{0.45\textwidth}
    \includegraphics[width=1\textwidth]{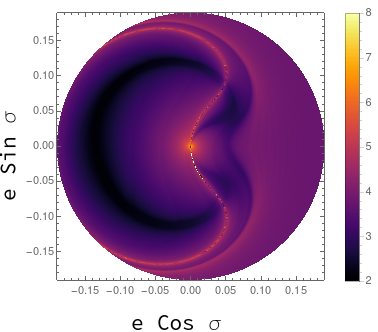}
    \end{subfigure}
    \caption{Replotting of the FLI maps of Figure \ref{smallecc} in the ($e \cos{\sigma},e \sin{\sigma}$) plane. The resulting plots are reminiscent of the ones in \cite{henlem}. The bright curve passing through the point (0,0) highlights the singularities of the action angle cylindrical formulation.}
    \label{smallecc_SFM}
\end{figure}
The maps in the last row of Figure \ref{smallecc} are bifurcation diagrams (obtained using FLIs) which depict the evolution of the equilibrium points, the sepatrices and of the $e=0$ level set under the variation of the area-to-mass ratio up to a value of $6$ m\textsuperscript{2}/kg. We remark that increasing the value of the area-to-mass ratio reduces the total number of equilibrium points from three to one. This result suggests that for small values of the eccentricity, the problem of SRP semi-secular resonances could be expressed in terms of the Second Fundamental Model of resonance, \cite{henlem}, or by its extension, the Extended Fundamental Model of resonance, \cite{breiter_ext}. To give further evidence in favour of this claim, in Figure \ref{smallecc_SFM} we replotted the FLI maps in the ($e\cos{\sigma},e\sin{\sigma}$)-plane. The resulting plots are similar to the ones presented in \cite{henlem}. In principle if one considers a maximum point of a resonant curve (expressed in the canonical variables) that is very close to $\eta=1$ (or $e=0$), one is going to have a similar situation with 2 PLEs, for a total of five equilibrium points, suggesting that, in this situation, the EFM by \cite{breiter_ext} could be more accurate for qualitatively describing the dynamics. Further investigations are necessary and they are going to be the subject of a future paper.

\subsection{Overlapping of resonances with $k=1$ at $i=90^\circ$}\label{overlappi}

\begin{figure}
    \centering
    \begin{subfigure}{0.45\textwidth}
    \includegraphics[width=\textwidth]{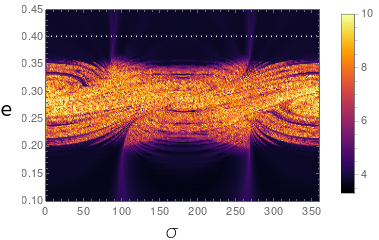}
    \end{subfigure}
    \begin{subfigure}{0.45\textwidth}
    \includegraphics[width=\textwidth]{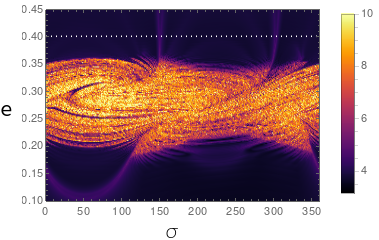}
    \end{subfigure}
    
    \begin{subfigure}{0.45\textwidth}
    \includegraphics[width=\textwidth]{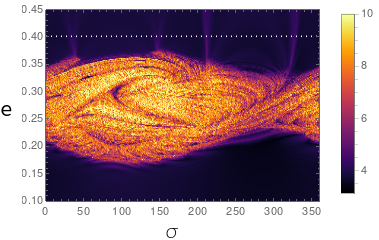}
    \end{subfigure}
    \begin{subfigure}{0.45\textwidth}
    \includegraphics[width=\textwidth]{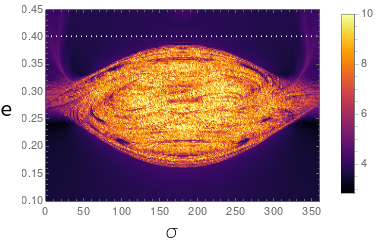}
    \end{subfigure}
    \caption{FLI study of the overlapping of resonances with $k=1$ for polar orbits. The maps are obtained by fixing the initial conditions $a=1.67 \ R_E, i=90^\circ, \ A/m=1$ m\textsuperscript{2}/kg and $\Om=\Om_0$ and computing the FLIs over a grid on the $(e,\sigma_{0,1})$-plane. The top plots correspond to $\Om_0=0^\circ$ and $\Om_0=45^\circ$, while the bottom ones correspond to $\Om_0=90^\circ$ and $\Om_0=180^\circ$. The dotted white line corresponds to the critical value of the eccentricity, equal to $0.401$.}
    \label{overlap}
\end{figure}

Finally, we discuss the overlapping of different resonant terms. 
In Appendix \ref{appC} we present several FLI maps on the ($i,e$)-plane which show the location of the stable and unstable equilibrium points together with the associated separatrices for a collection of values of the semi-major axis and of the area-to-mass ratio. In particular, in Figures \ref{comp10} - \ref{comp10180} one can appreciate how the resonant curves overlap for $A/m= 1$ and $10$ m\textsuperscript{2}/kg. From these figures we deduce that for moderate values of the semi-major axis $a$ there could be many intersections between the resonant curves associated to different resonant terms. Moreover, note that the FLI maps exhibit a strong chaotic behaviour in the neighborhood of an intersection between two (or more) resonant curves. 

We focus on the overlapping of the resonances $(0,1)$, $(1,1)$ and $(-1,1)$ at $i=90^\circ$, i.e. for polar orbits. Since an analytical description of this problem is beyond the scope of this article, we  proceed numerically using FLIs. The results are described in terms of the usual orbital elements since by considering more than one resonant term, the quantities $\alpha_j$ are no longer constants of motion. The phase portrait is presented using a FLI map on the $(e,\sigma_{0,1})$-plane, by fixing all the other initial conditions in terms of the usual orbital elements. Since we are using $\sigma_{0,1}$ as the resonant angle, if the object has a nonzero value of $\Om_0$ the locations of the equilibria induced by the resonances $(\pm 1,1)$ are shifted left or right, in opposite directions, resulting in a total shift in the two centers of $2 \Om_0$. On the other hand, the location of the equilibrium points for the resonance $(0,1)$ is the same (PLE with the stable point at $\sigma_{0,1}=180^\circ$). In Figure \ref{overlap} we show the phase portraits for four different values of $\Om_0$, which is nearly constant in vision of Eq. (\ref{roc2}). Depending on its value the qualitative picture changes, ranging from a chaotic entanglement to a slightly more regular structure where one can distinguish the stable resonant island of the resonance $(0,1)$  filled with chaotic orbits.

The above examples are generated considering $a=1.67\ R_E$, $i=90^\circ$, $A/m=1$ m\textsuperscript{2}/kg. This overlapping affects polar orbits with $a$ ranging from $1.589\ R_E$ to $1.80\  R_E$, and at the lower altitudes it could result in a chaotic variation of the eccentricity, which might produce a forced re-entry if the eccentricity reaches the critical limit. Mission designers should avoid this "belt" for the long-term missions.

\section{Conclusions, Final Remarks and Future Work}\label{concl}
The location of the equilibria and their stability for SRP semi-secular resonances has been studied by adapting the procedure described in \cite{breiter_aps} for the analysis of lunisolar apsidal resonances. A set of formulas for estimating the amplitude of the resonant islands surrounding the stable equilibria was derived and applied to identify the strongest SRP resonances. From the analytical formulas it stems that the amplitude of the resonant islands is proportional to the square root of the area-to-mass ratio, while the period is inversely proportional to the same quantity. The analytical formulas were validated by extensive numerical testing using the Fast Lyapunov Indicators, together with Hamiltonian and Cartesian propagators. Each resonance was investigated using a toy model to describe the influence of a specific term on the overall dynamics. We described the phenomenon of merging of nearby resonances and the behaviour close to $e=0$ (where virtual singularities appear). Similarities with the Extended Fundamental Model (EFM) of resonance have been highlighted. We conjecture that the problem of SRP semi-secular resonances can be modeled using the EFM after removing the virtual singularity using \textit{Poincaré-like elements}. Finally, we focused on the numerical study of the overlapping of three SRP resonances for polar orbits in low orbit.

In practice, SRP semi-secular resonances appear to be extremely relevant for objects in the high-LEO/low-MEO region with moderately high values of the area-to-mass ratio. In particular, in Section \ref{FLIsec}, we showed some examples which highlight how an object with low eccentricity could experience a large variation in the perigee distance, for given values of the area-to-mass ratio and of the integrals of motion of the problem. If the area-to-mass ratio is too large, or if one equilibrium point is too close to the $e=0$ line, the debris could experience some large perigee variations which might even result in a collision with the planet if left uncontrolled. 

\color{black}
Future space missions involving a solar sail could benefit from the results of this paper, thanks to the formulas from Section \ref{model}, which provide immediate insights on the maximum variations in the orbital elements and the timescales upon which SRP resonances act. Mission designers could for example exploit the resonance $(1,-1)$ to produce large variations in the perigee in the span of a few years or even months, depending on the size of the equipped solar sails without the need for using external propulsion, thus improving the efficiency of the mission.

Finally, in view of the results of this paper, when designing missions for satellites with large area-to-mass ratio one should avoid the region corresponding to secular lunisolar resonances, polar orbits, and generically speaking, regions where two SRP resonances overlap, since, depending on $t_0$ and on the initial conditions of the object, the chaotic overlapping of the SRP semi-secular resonances might increase the eccentricity up to the critical value. 
\color{black}
Future works which stem from this research include (but are not limited to): the removal of the virtual singularities to prove that SRP semi-secular resonances can indeed be modeled using either the EFM or the SFM; the study of the overlapping of two different SRP resonant terms; the study of the overlapping of second-degree lunisolar semi-secular resonances and the first-degree SRP ones; the formal study of the phenomenon of the merging of resonances.

\subsection*{Acknowledgements}
\noindent The Author would like to thank Prof. C\u{a}t\u{a}lin Gale\c{s}, Prof. Christos Efthymiopoulos and Edoardo Legnaro for the useful discussions during the writing of this manuscript, and  for providing many useful references, and two anonymous reviewers for their insightful suggestions which have improved the quality of this paper.
\appendix
\section{Series expansion of the perturbations}\label{appA}
We collect some classical expressions for the most relevant perturbations of the two-body problem. They are presented using the usual Keplerian osculating elements and can be treated as Hamiltonian functions after converting to Delaunay elements.
\subsection{The perturbation due to the Earth}
In a geocentic quasi-inertial frame, the Hamiltonian term $\mathcal{H}_{geo}$ can be written as
\begin{equation}\label{hea}
    \mathcal{H}_{geo}=-\dfrac{\mu_E}{a}\sum_{n=2}^{\infty}\sum_{m=0}^{n}\left(\dfrac{R_E}{a}\right)^n\sum_{p=0}^{n}F_{nmp}(i)\sum_{q=-\infty}^{\infty}G_{npq}(e)S_{nmpq}(M,\omega,\Omega,\theta).
\end{equation}
The functions $F_{nmp}(i)$ are called \textit{Kaula inclinations functions} and are defined as
\begin{align}\label{kauf}
        F_{nmp}(i)=\sum _w& \dfrac{(2n-2w)!}{w!(n-w)!(n-m-2w)!2^{2n-2w}}\sin^{n-m-2w}i\sum_{s=0}^m\binom{m}{s}\cos^s i \nonumber\\
        &\times \sum _c \binom{n-m-2w+s}{c}\binom{m-s}{p-w-c}(-1)^{c-k},
\end{align}
where $k$ is the integer part of $\frac{n-m}{2}$, the index $w$ runs between zero and the minimum between $p$ and $k$, while $c$ is taken over all values which give nonzero binomial coefficients.
\smallskip

\noindent The \textit{eccentricity functions} $G_{npq}(e)$ are given by
\begin{equation}
    G_{npq}(e)=(-1)^{|q|}(1+\beta^2)^n \beta^{|q|} \sum_{k=0}^{infty} P_{npqk} Q_{npqk}\beta^{2k},
\end{equation}
where $$\beta=\dfrac{e}{1+\sqrt{1-e^2}},$$
while the functions $P_{npqk}$ and $Q_{npqk}$ are given by
$$
P_{npqk}=\sum_{r=0}^{h}\binom{2p'-2n}{h-r}\dfrac{(-1)^r}{r!}\left(\dfrac{(n-2p'+q')e}{2\beta}\right)^r,
$$
where $h=k+q'$ when $q'>0$ and $h=k-q'$ when $q'<0$, and
$$
Q_{npqk}=\sum_{r=0}^{h}\binom{-2p'}{h-r}\dfrac{1}{r!}\left(\dfrac{(n-2p'+q')e}{2\beta}\right)^r,
$$
where $h=k$ when $q'>0$ while $h=k-q'$ when $q'<0$, $p'=p$ and $q'=q$ when $p\leq \frac{n}{2}$, while $p'=n-p$ and $q'=-q$ when $p>\frac{n}{2}$.

\noindent The quantities $S_{nmpq}$ in (\ref{hea}) are defined by
$$
S_{nmpq}=\left[\begin{smallmatrix} C_{nm} \\ -S_{nm} \end{smallmatrix}\right]_{n-m \ odd}^{n-m \ even}\cos \psi_{nmpq} + \left[\begin{smallmatrix} S_{nm} \\ C_{nm} \end{smallmatrix}\right]_{n-m \ odd}^{n-m \ even}\sin \psi_{nmpq},
$$
where 
$$
\psi_{nmpq}=(n-2p)\omega + (n-2p+q)M + m(\Omega-\theta).
$$
\noindent See \cite{kaula_spher} for the derivation of the above functions.
\subsection{The lunisolar perturbations}

We first focus on the gravitational potential due to the Sun. Let us assume that the Sun moves on a Keplerian orbit with semi-major axis $a_S=1$ au , eccentricity $e_S=0.0167$, inclination $i_S=23^\circ 26' 21.406''$, argument of perigee $\omega_S=282.94^\circ$, longitude of the ascending node $\Omega_S=0^\circ$. The rate of change of the mean anomaly is $\dot{M_S}\simeq1^\circ/\text{day}$. The solar orbital elements $\mathcal{Y}_S=(a_S,e_S,i_S,M_S,\omega_S,\Omega_S)$ are referred to the celestial equator, exactly as the debris orbital elements $\mathcal{Y}=(a,e,i,M,\omega,\Omega)$.
In \cite{kaulun62} the solar disturbing function $\mathcal{R}_{S}$ is given by\footnote{One has that $\H_{S}=-\mathcal{R}_S$.}
\begin{align}\label{rsun}
    \mathcal{R}_{S}=&\mathcal{G}m_S\sum_{l=2}^{\infty}\sum_{m=0}^{l}\sum_{p=0}^{l}\sum_{h=0}^{l}\sum_{q=-\infty}^{\infty}\sum_{j=-\infty}^{\infty} \dfrac{a^l}{a_S^{l+1}}\epsilon_m \dfrac{(l-m)!}{(l+m)!} \nonumber \\
    &\times\mathcal{F}_{lmph}(i,i_S)\mathcal{H}_{lpq}(e)\mathcal{G}_{lhj}(e_S)\cos(\varphi_{lmphqj})
\end{align}
where 
\begin{align*}
    \mathcal{F}_{lmph}&\equiv  F_{lmp}(i)F_{lmh}(i_S), \\
    \varphi_{lmphqj} &\equiv (l-2p)\omega + (l-2p+q)M-(l-2h)\omega_S-(l-2h+j)M_S + m(\Omega -\Omega_S),
\end{align*}
and the functions $F_{lmp}(i)$ and $F_{lmh}(i_S)$ are the same Kaula inclination functions defined when discussing the geopotential expansion. Moreover, $m_S$ denotes the mass of the Sun and the quantities $\epsilon_m$ are defined as
\begin{equation*}
    \epsilon_m=\begin{cases}
    1 \text{  if } m=0, \\
    2\text{  if } m\in \Z\setminus\{0\}.
    
    \end{cases}
\end{equation*}
Finally, the functions $\mathcal{H}_{lpq}(e)$ and $G_{lhj}(e_S)$ correspond to the Hansen coefficients $X_{l-2p+q}^{l,l-2p}(e)$ and $X_{l-2h+j}^{-(l+1),l-2h}(e_S)$, respectively. For more details about the Hansen coefficients we refer the reader to \cite{hansen}.
\medskip

We now focus on the lunar disturbing function. In order to describe the motion of the Moon we adopt Keplerian elements with respect to the ecliptic plane. As a consequence, the inclination $i_M$ becomes nearly constant, while the changes in the argument of perigee $\omega_M$ and in the longitude of the ascending node $\Omega_M$ become approximately linear with respect to time, with rates of change equal to, respectively, $\dot{\omega}_M\simeq 0.164 ^\circ/\text{day}$ and $\dot{\Omega}_M\simeq-0.053^\circ/\text{day}$. Moreover, the mean anomaly changes as $\dot{M}_M\simeq13.06^\circ /\text{day}$.
We also assume that the Moon moves on a Keplerian ellipse with semi-major axis $a_M=$ 384 748 km, eccentricity $e_M=0.0549$ and inclination (referred to the ecliptic) $i_M=5^\circ 15'$. For an accurate description of the orbital elements used for the Moon we refer the reader to \cite{celros} and \cite{celletti17}.
\smallskip

The lunar disturbing Hamiltonian is given by\footnote{As in the solar case, one has that $\H_M=-\mathcal{R}_M$}
\begin{align}\label{rmoon}
    \mathcal{R}_{M}=&\ \mathcal{G}m_M\sum_{l\geq2}\sum_{m=0}^l\sum_{p=0}^l\sum_{s=0}^l\sum_{q=0}^l\sum_{j=-\infty}^{\infty}\sum_{r=-\infty}^{\infty}(-1)^{m+s}\nonumber \\
    &\times (-1)^{k_1}\dfrac{\epsilon_M\epsilon_S}{2a_M}\dfrac{(l-s)!}{(l+m)!}\left( \dfrac{a}{a_M}\right)^l F_{lmp}(i)F_{lsq}(i_M)\mathcal{H}_{lpj}(e)\mathcal{G}_{lqr}(e_M) \nonumber\\
    &\times \{ (-1)^{k_2}U_{l}^{m,-s}\cos(\Bar{\theta}_{lsqr}+\Bar{\theta}'_{lmpj}-y_s\pi) \nonumber \\
    & + (-1)^{k_3}U_{l}^{m,s}\cos(\Bar{\theta}_{lmpj}-\Bar{\theta}'_{lsqr}-y_s\pi)\},
\end{align}
where $y_s=0$ for $s$ even and $y_s=\frac{1}{2}$ when $s$ is odd, $k_1=\left[\frac{m}{2}\right]$, $k_2=t(m+s-1) + 1$, $k_3=t(m+s)$ with $t=(l-1) \text{ mod 2}$; the quantities $\Bar{\theta}_{lmpj}$ and $\Bar{\theta}'_{lsqr}$ are given by
\begin{align*}
    \Bar{\theta}_{lmpj}=& (l-2p)\omega + (l-2p + j)M+m\Omega \\
    \Bar{\theta}'_{lsqr}=&(l-2q)\omega_M + (l-2q + r)M_M+s\left(\Omega_M-\frac{\pi}{2}\right); 
\end{align*}
the functions $U_l^{m,s}$ are defined as
\begin{align*}
    U_l^{m,s}=& \sum_{r=\max (0,-(m+s))}^{\min(l-s,l-m)}(-1)^{l-m-r}\binom{l+m}{m+s+r}\binom{l-m}{r}\cos^{m+s+2r}\left(\dfrac{\varepsilon}{2}\right)\\
    &\sin^{-m-s+2(l-r)}\left(\dfrac{\varepsilon}{2}\right)
    \end{align*}
where $\varepsilon$ denotes the obliquity of the ecliptic, which is $\varepsilon=23^\circ26'21.45''$. The Kaula inclination functions and the eccentricity functions corresponding to the Hansen coefficients are the same as in the solar case.
\smallskip

\noindent We refer the interested reader to \cite{kaula_spher} and  \cite{celletti17} for a thorough exposition of the derivation of the above expansions.

\subsection{Solar Radiation Pressure perturbation}\label{chapsrp}
The expansion of the SRP potential (\cite{hug77}) is basically the same as for the solar one, except for the coefficient $\mathcal{G} m_S$ which is replaced by $ C_r P_r \frac{A}{m}$, and for the inclusion of the first degree terms, i.e. those with $l=1$:
\begin{align}\label{hsrp}
 \mathcal{H}_{SRP}= \ & C_r P_r \dfrac{A}{m}\sum_{l=1}^{\infty}\sum_{m=0}^{l}\sum_{p=0}^{l}\sum_{h=0}^{l}\sum_{q=-\infty}^{\infty}\sum_{j=-\infty}^{\infty}\dfrac{a^l}{a_S^{l-1}}\epsilon_m
\dfrac{(l-m)!}{(l+m)!}\nonumber\\
&\times \mathcal{F}_{lmph}\left(i,i_S\right)\mathcal{H}_{lpq}(e)\mathcal{G}_{lhj}(e_S)\cos(\varphi_{lmphqj}),
\end{align}
where all functions and numbers appearing in the formula above have already been defined in the previous sections. \cite{hug77} provides a detailed explanation on how to detect the most relevant terms in the SRP expansion, and lists the six first-degree terms appearing in this paper as the ones of greatest magnitude.

\section{Fast Lyapunov Indicators}\label{appB}
Fast Lyapunov Indicators (FLIs) are \textit{chaos indicators} used to numerically investigate the stability of a dynamical system and are strongly related to the Lyapunov Charateristic Exponents\footnote{FLIs are basically the value of the largest Lyapunov characteristic exponent at a \textit{fixed time}. }. Comparing the values of the FLIs as the initial conditions of parameters are varied, one can distinguish between regular, resonant or chaotic motions. Here we briefly recall the definition of FLIs. Let us consider the problem defined by the vector field $f$
\begin{equation*}
    \dot{\xi}=f(\xi), \quad \xi \in \mathbb{R}^6
\end{equation*}
and its corresponding \textit{variational equations}, which describe the evolution of a vector $\eta$ on the tangent space
\begin{equation*}
    \dot{\eta}=\left(\dfrac{\de f(\xi)}{\de \xi}\right)\eta, \quad \eta\in T_{\xi}\mathbb{R}^6\equiv\mathbb{R}^6
\end{equation*}
The explicit computation of the FLI proceeds as follows: the FLI at a given time $T\geq 0$ is obtained by the expression
\begin{equation*}
    \text{FLI}(\xi(0),\eta(0),T)\equiv \sup_{0<t\leq T}\log\|\eta(t)\|,
\end{equation*}
where $\xi(0)$ and $\eta(0)$ are the initial conditions for the above problem and its variational equations, while $\eta(t)$ is the solution of the variational equation at time $t$ and $\|.\|$ is a suitably chosen norm on $\mathbb{R}^6$. Practically speaking, one fixes an initial tangent vector and a suitably chosen reference time $T$. Then, one proceeds to numerically solve the problem and its variational equations to compute the FLIs, which can be stored in a data file and then pictured using a colour map. The choice of the reference time $T$ and of the initial tangent vector depend on the problem at hand. For more information, we refer the reader to \cite{guzzo} and \cite{benettin}. 

\section{Location of the equilibria in the $(i,e)$ plane}\label{appC}
\begin{figure}[ht!]
\centering

\begin{subfigure}{.3\textwidth}
    \centering
    \includegraphics[width=0.95\textwidth]{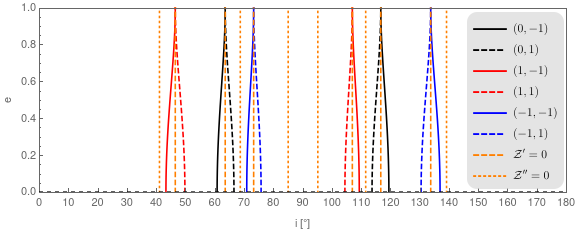}
    \subcaption{\tiny $a\leq7403.31$} 
\end{subfigure}  \begin{subfigure}{.3\textwidth}
    \centering
    \includegraphics[width=0.95\textwidth]{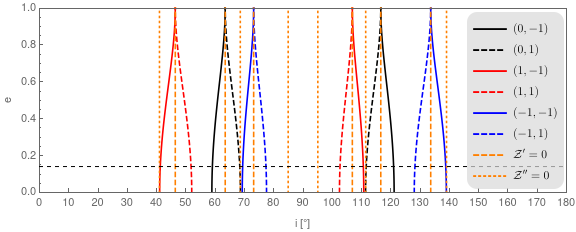}
    \subcaption{\tiny $7403.31<a\leq 7445.06$} 
\end{subfigure} \begin{subfigure}{.3\textwidth}
    \centering
    \includegraphics[width=0.95\textwidth]{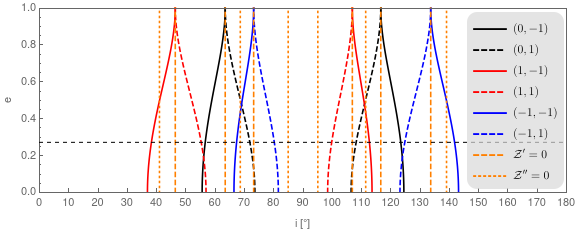}
    \subcaption{\tiny$7445.06<a\leq 9453.98$} 
\end{subfigure} 

\begin{subfigure}{.3\textwidth}
    \centering
    \includegraphics[width=0.95\textwidth]{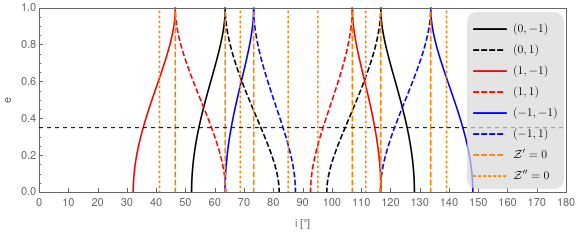}
    \subcaption{\tiny $9453.98<a\leq 10133.2$} 
\end{subfigure} \begin{subfigure}{.3\textwidth}
    \centering
    \includegraphics[width=0.95\textwidth]{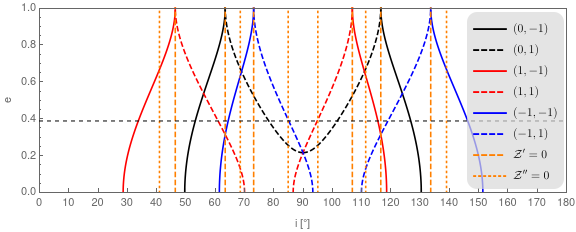}
    \subcaption{\tiny$10133.2<a\leq 10675$} 
\end{subfigure}
\begin{subfigure}{.3\textwidth}
    \centering
    \includegraphics[width=0.95\textwidth]{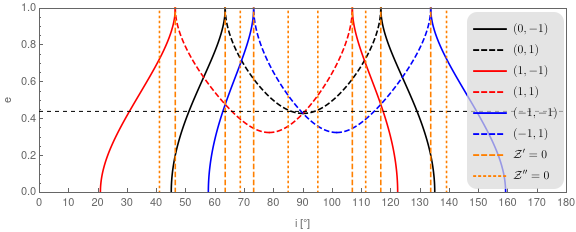}
    \subcaption{\tiny$10675<a\leq 12352.5$} 
\end{subfigure}

\begin{subfigure}{.3\textwidth} 
    \centering
    \includegraphics[width=0.95\textwidth]{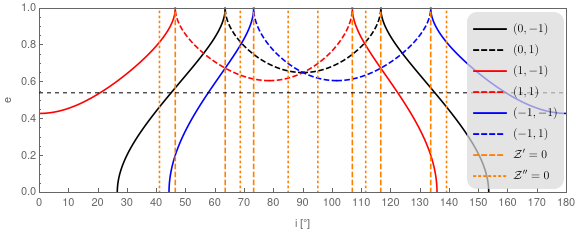}
    \subcaption{\tiny$12352.5<a\leq 15057.9$} 
\end{subfigure}
\begin{subfigure}{.3\textwidth}
    \centering
    \includegraphics[width=0.95\textwidth]{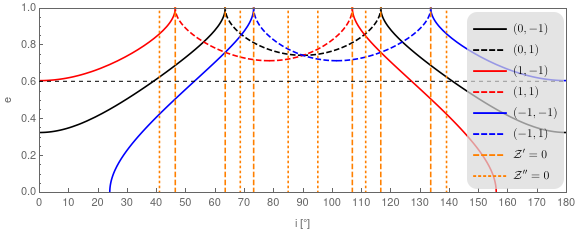}
    \subcaption{\tiny$15057.9<a\leq 16907.3$} 
\end{subfigure} \begin{subfigure}{.3\textwidth}
    \centering
    \includegraphics[width=0.95\textwidth]{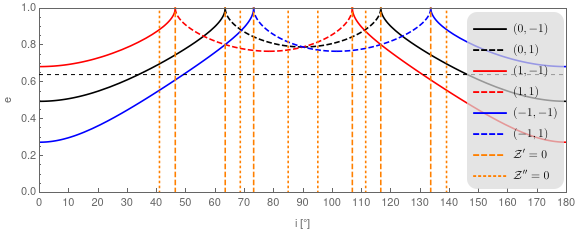}
    \subcaption{\tiny$a> 16907.3$}  
\end{subfigure}
    \caption{Approximate location of SRP semi-secular resonances in the $(i,e)$-plane. The actual location of the stable and ustable equilibrium points is close to the predicted one if $\frac{A}{m}$ is small enough and we are far enough from critical regions.}
    \label{allie}
\end{figure}

\begin{figure}[ht!]
    \centering
    \includegraphics[width=0.98\textwidth]{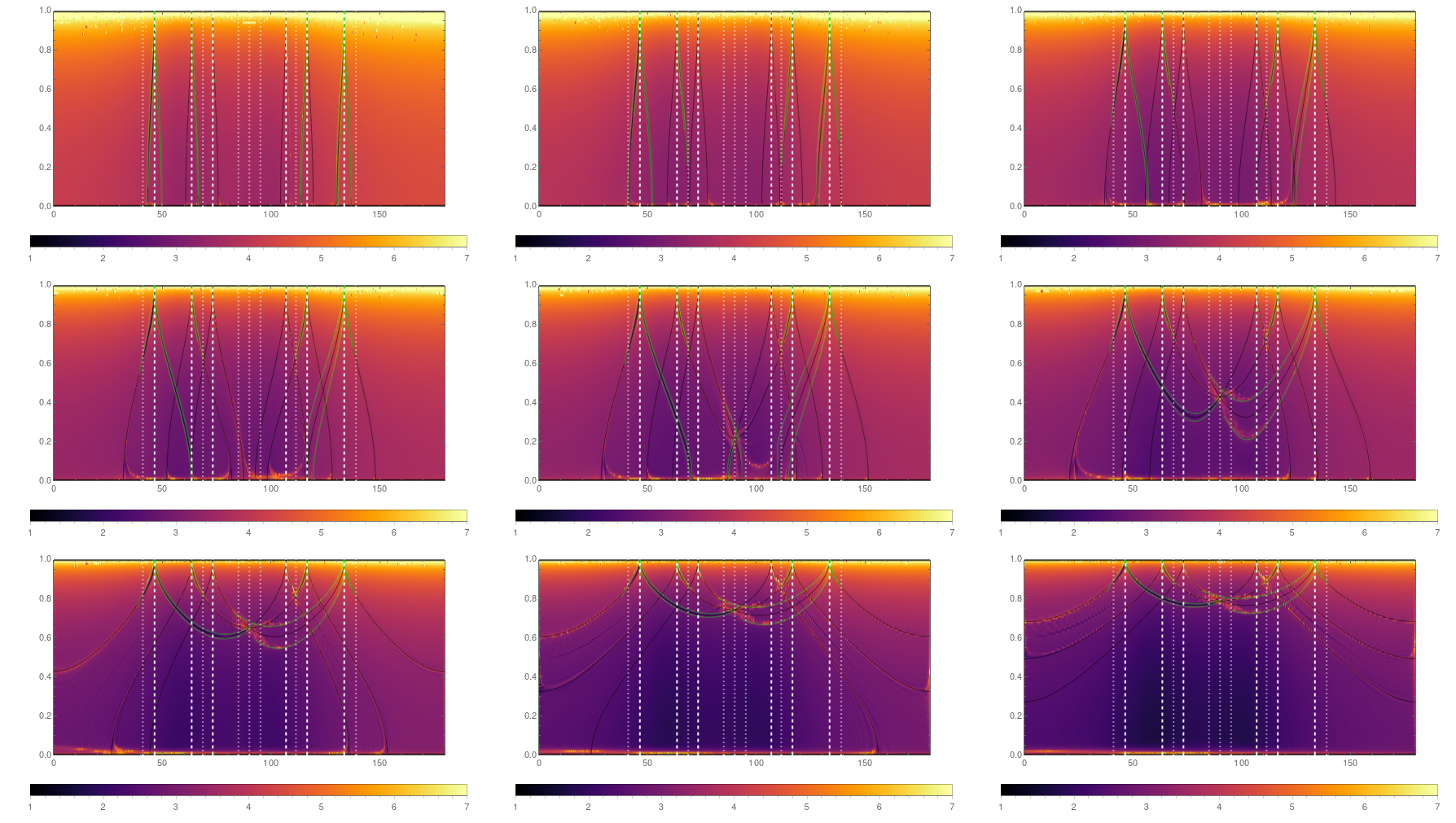}
    \caption{\footnotesize FLI maps in the ($i,e$) plane, depicting the location of the stable points (\textit{dark colors}), unstable points and separatrices (\textit{bright colors}), for an area-to-mass ratio of $1$ m\textsuperscript{2}/kg , with $\Om(t_0)=0^\circ$. The initial time is chosen so that $M_S(t_0)+\om_S=0$. The nine maps correspond to the ones presented in Figure \ref{allie}. $\om(t_0)$ is chosen so that $\sigma_{j,k}=0^\circ$. \newline
    The green curves represent the positions of the separatrices of the integrable resonant approximation $\overline{\mathcal{K}}$ given by Eq. (\ref{apppend}).}
    \label{comp10}
\end{figure}

\begin{figure}[ht!]
    \centering
    \includegraphics[width=0.98\textwidth]{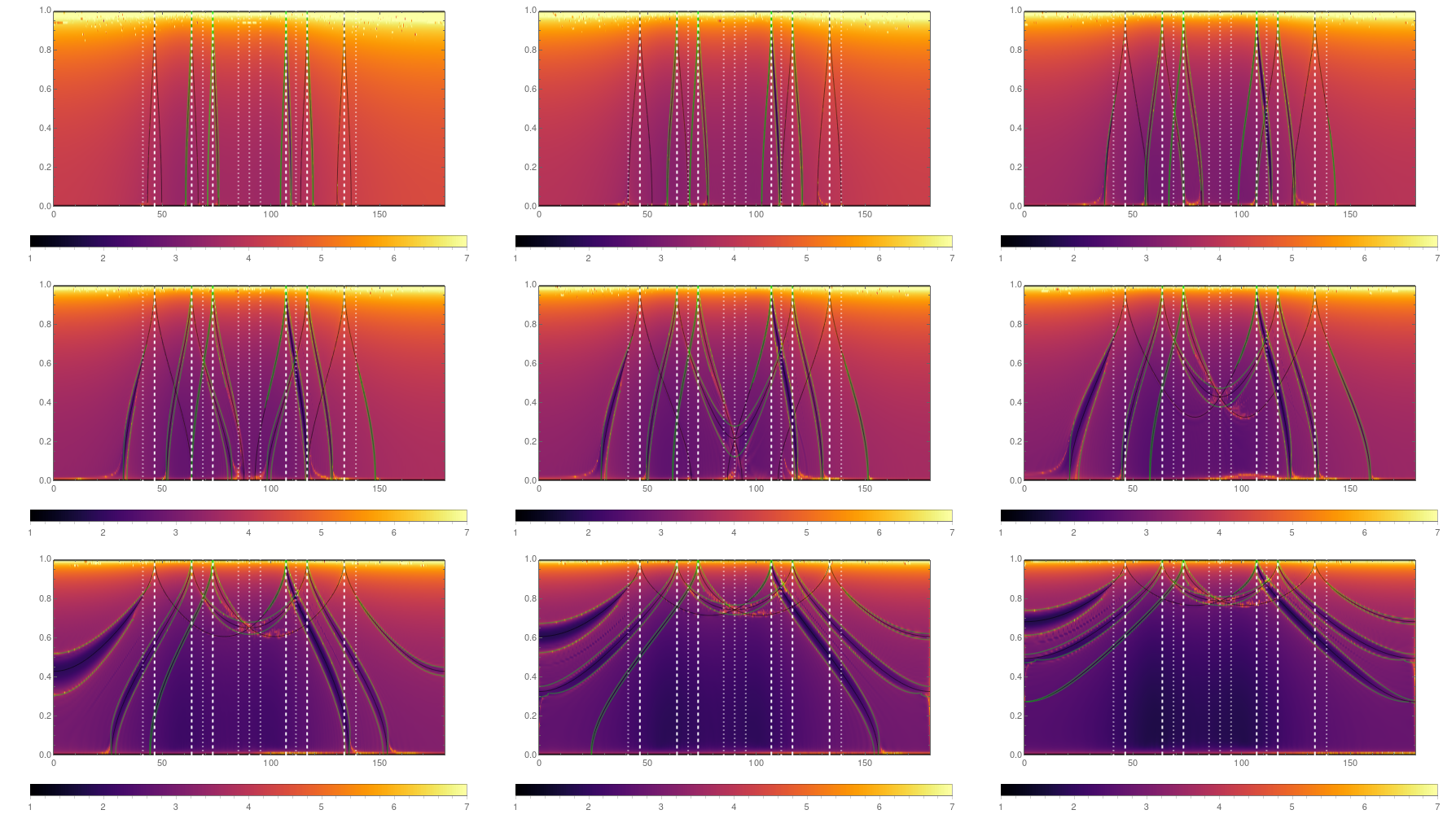}
    \caption{\footnotesize FLI maps in the ($i,e$) plane, depicting the location of the stable points (\textit{dark colors}), unstable points and separatrices (\textit{bright colors}), for an area-to-mass ratio of $1$ m\textsuperscript{2}/kg , with $\Om(t_0)=0^\circ$. The initial time is chosen so that $M_S(t_0)+\om_S=0$. The nine maps correspond to the ones presented in Figure \ref{allie}. $\om(t_0)$ is chosen so that $\sigma_{j,k}=180^\circ$. \newline
    The green curves represent the positions of the separatrices of the integrable resonant approximation $\overline{\mathcal{K}}$ given by Eq. (\ref{apppend}).}
    \label{comp1180}
\end{figure}

\begin{figure}[ht!]
    \centering
    \includegraphics[width=0.98\textwidth]{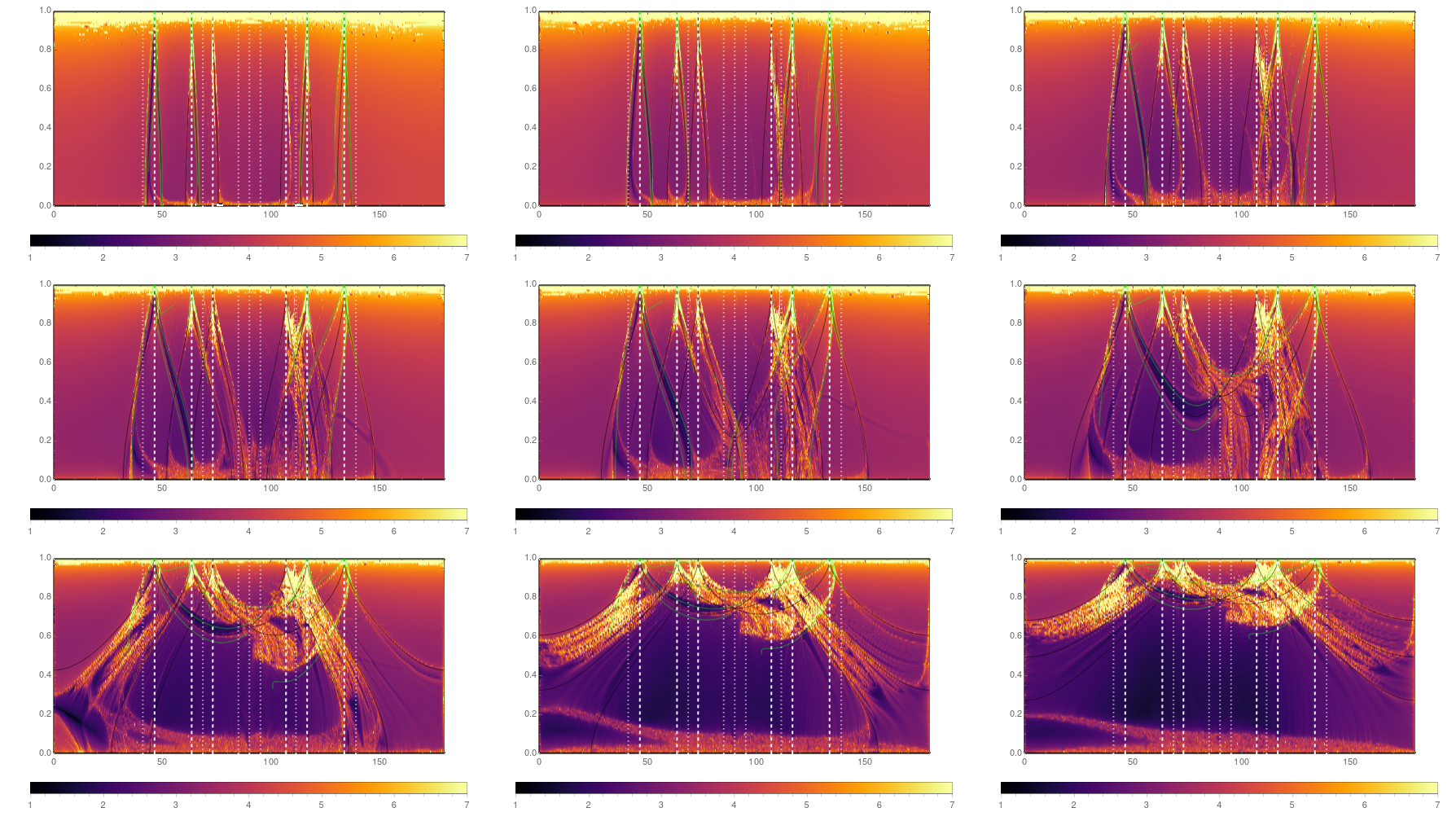}
    \caption{\footnotesize FLI maps in the ($i,e$) plane, depicting the location of the stable points (\textit{dark colors}), unstable points and separatrices (\textit{bright colors}), for an area-to-mass ratio of $10$ m\textsuperscript{2}/kg , with $\Om(t_0)=0^\circ$. The initial time is chosen so that $M_S(t_0)+\om_S=0$. The nine maps correspond to the ones presented in Figure \ref{allie}. $\om(t_0)$ is chosen so that $\sigma_{j,k}=0^\circ$. \newline
    The green curves represent the positions of the separatrices of the integrable resonant approximation $\overline{\mathcal{K}}$ given by Eq. (\ref{apppend}).}
    \label{comp100}
\end{figure}

\begin{figure}[ht!]
    \centering
        \includegraphics[width=0.98\textwidth]{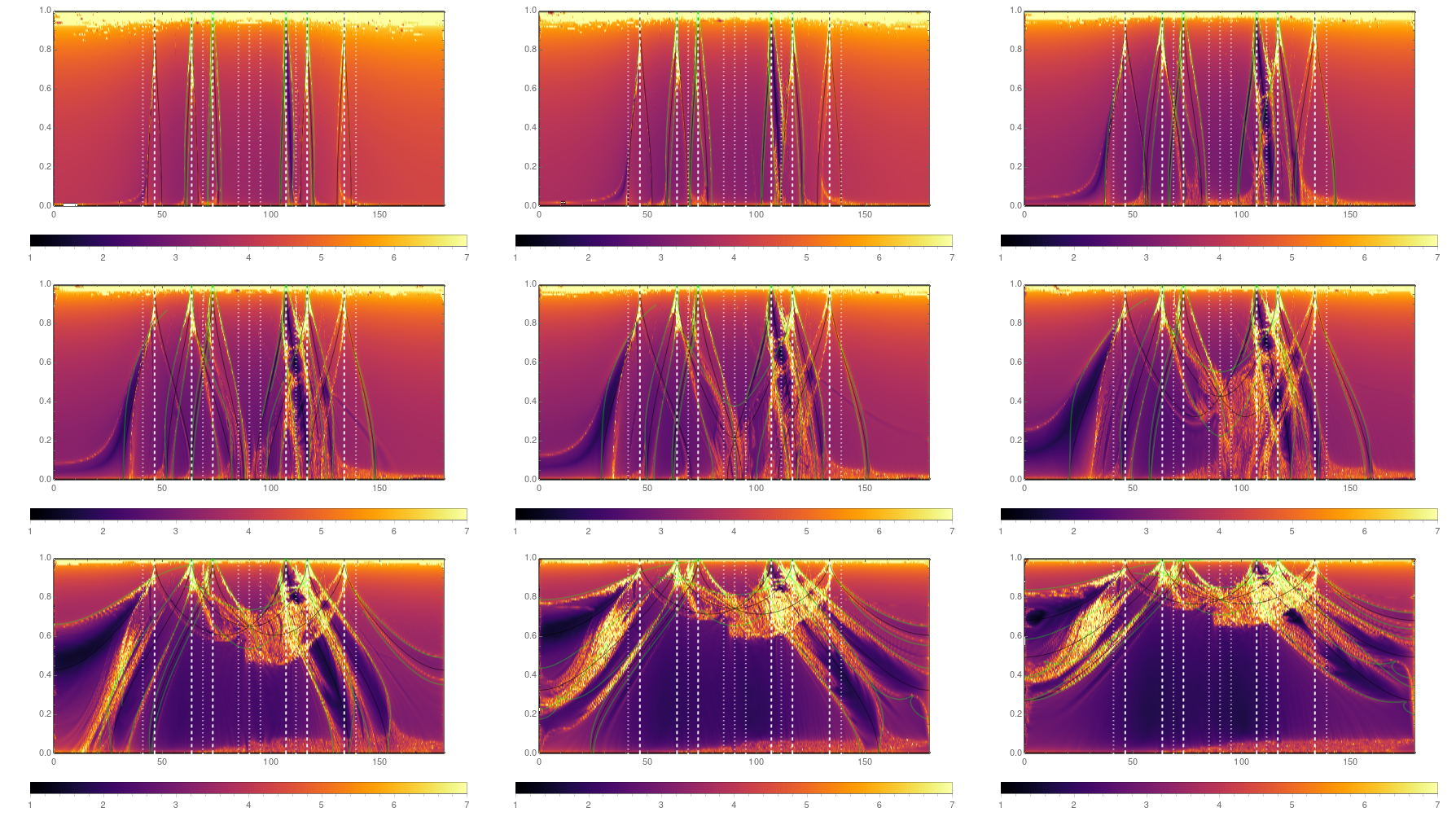}
    \caption{\footnotesize FLI maps in the ($i,e$) plane, depicting the location of the stable points (\textit{dark colors}), unstable points and separatrices (\textit{bright colors}), for an area-to-mass ratio of $10$ m\textsuperscript{2}/kg , with $\Om(t_0)=0^\circ$. The initial time is chosen so that $M_S(t_0)+\om_S=0$. The nine maps correspond to the ones presented in Figure \ref{allie}. $\om(t_0)$ is chosen so that $\sigma_{j,k}=180^\circ$. \newline
    The green curves represent the positions of the separatrices of the integrable resonant approximation $\overline{\mathcal{K}}$ given by Eq. (\ref{apppend}).}
    \label{comp10180}
\end{figure}

\begin{figure}[ht!]
    \centering
    \includegraphics[width=0.98\textwidth]{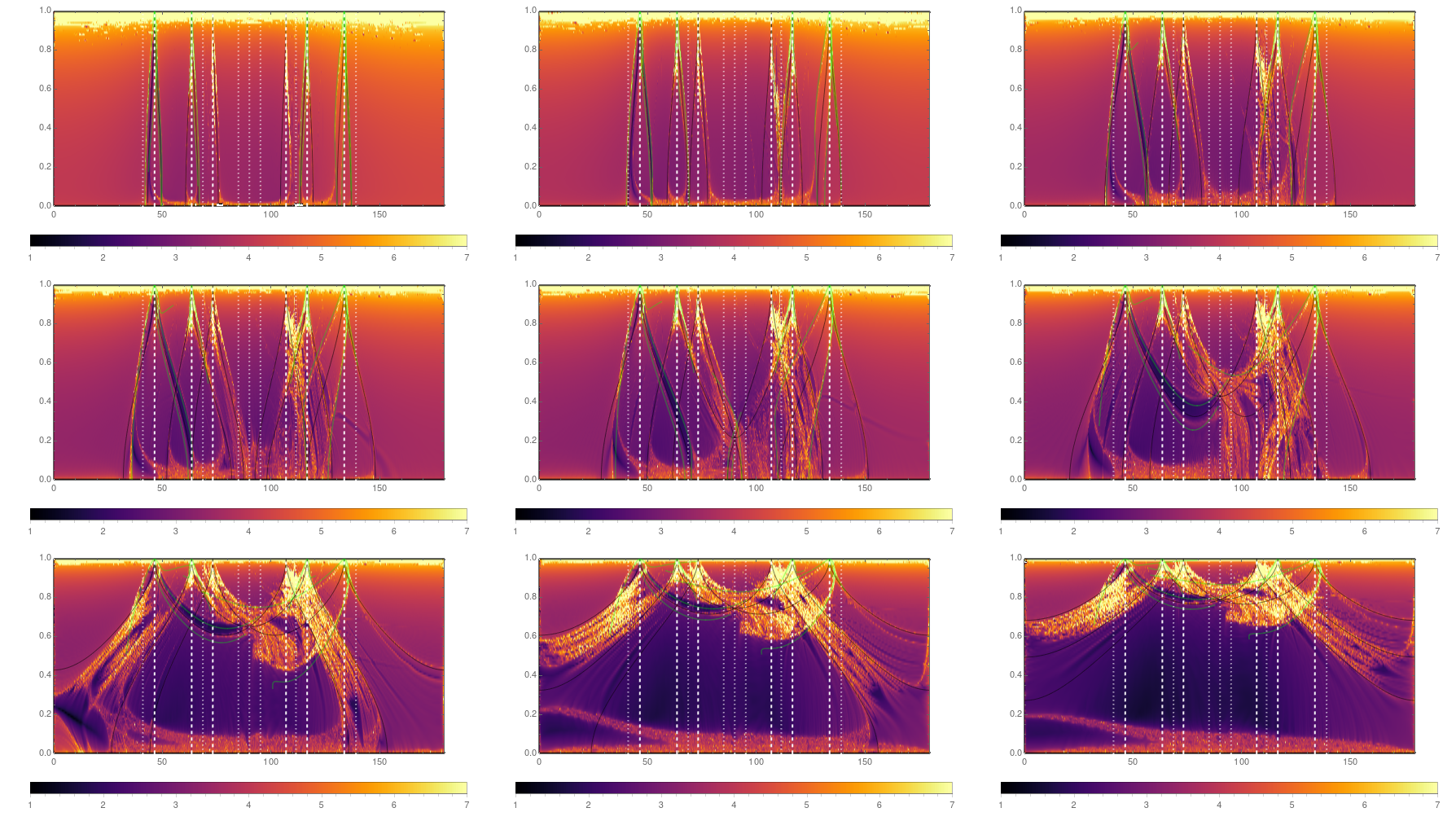}
    \caption{\footnotesize FLI maps in the ($i,e$) plane, depicting the location of the stable points (\textit{dark colors}), unstable points and separatrices (\textit{bright colors}), for an area-to-mass ratio of $10$ m\textsuperscript{2}/kg, with $\Om(t_0)=30^\circ$. The initial time is chosen so that $M_S(t_0)+\om_S=0$. The nine maps correspond to the ones presented in Figure \ref{allie}. \newline
    The green curves represent the positions of the separatrices of the integrable resonant approximation $\overline{\mathcal{K}}$ given by Eq. (\ref{apppend}).}
    \label{comp100om30}
\end{figure}

\begin{figure}[ht!]
    \centering
    \includegraphics[width=0.98\textwidth]{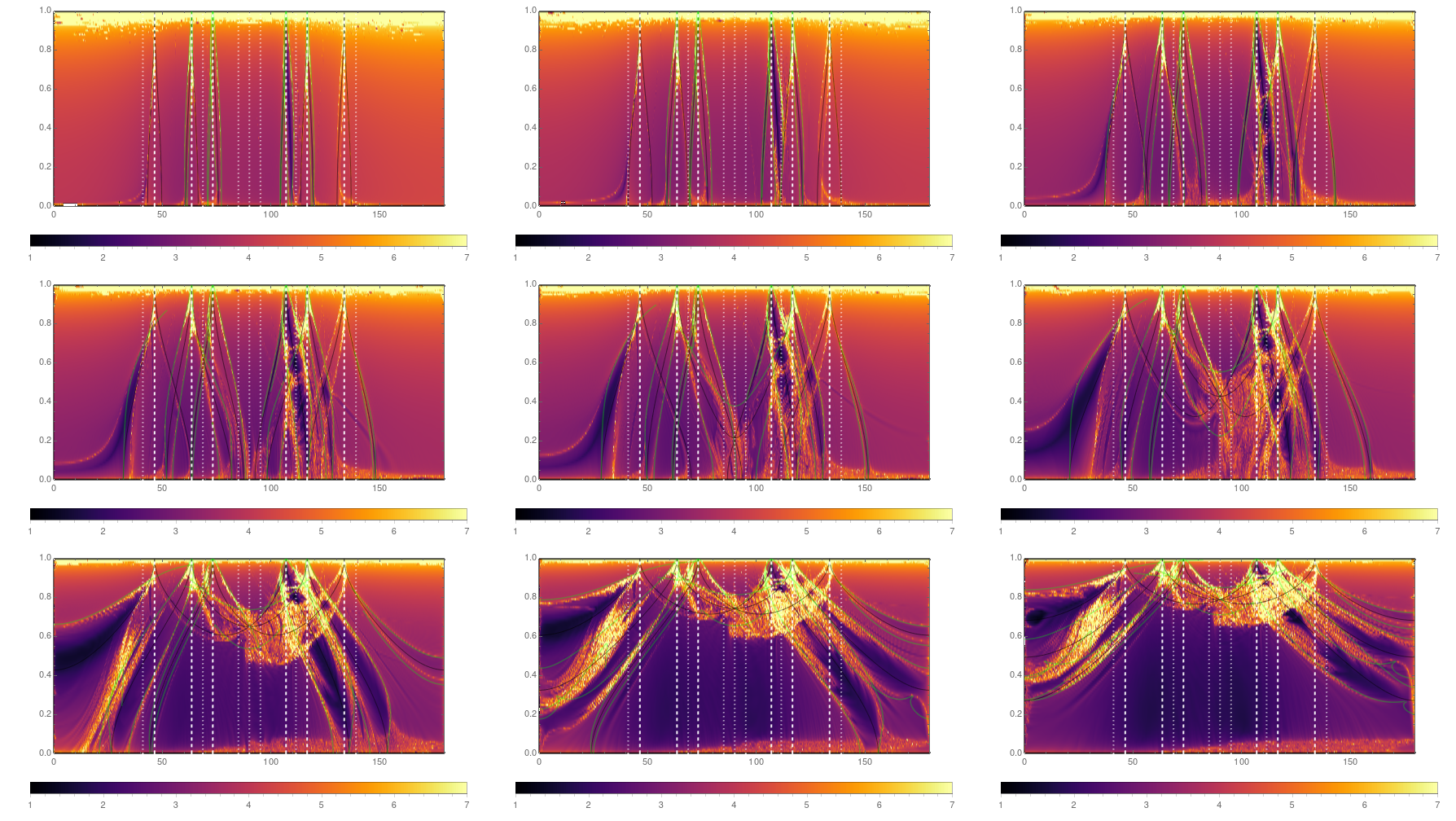}
    \caption{\footnotesize FLI maps in the ($i,e$) plane, depicting the location of the stable points (\textit{dark colors}), unstable points and separatrices (\textit{bright colors}), for an area-to-mass ratio of $10$ m\textsuperscript{2}/kg, with $\Om(t_0)=30^\circ$. The initial time is chosen so that $M_S(t_0)+\om_S=0$. The nine maps correspond to the ones presented in Figure \ref{allie}. \newline
    The green curves represent the positions of the separatrices of the integrable resonant approximation $\overline{\mathcal{K}}$ given by Eq. (\ref{apppend}).}
    \label{comp10180om30}
\end{figure}

Here we collect some plots showing the location of resonances and their associated resonant equilibria in the $(i,e)$ plane for various values of the semi-major axis, which represent all possible qualitatively different configurations of the resonant curves described in Section \ref{approxcha}. In particular, we are going to consider the following values for $z_p$, or equivalently $a$:
\begin{itemize}
    \item $z_p=0.2$, or $a=6397.94$ km\footnote{In this situation the maximum eccentricity is very small and in principle one should not disregard the effect of drag. Nonetheless, we included this case in order to provide a complete analysis.},
    \item $z_p=0.336$, or $a=7420.18$ km,
    \item $z_p=0.6$, or $a=8757.11$ km,
    \item $z_p=0.9$, or $a=9832.69$ km, 
    \item $z_p=1.1$, or $a=10412.9$ km,
    \item $z_p=1.5$, or $a=11377.8$ km,
    \item $z_p=3$, or $a=13869.7$ km,
    \item $z_p=5$, or $a=16049.1$ km,
    \item $z_p=7$, or $a=17668.6$ km.
\end{itemize}
Figure \ref{allie} is a collection of the resonant curves which where individually presented throughout Section \ref{approxcha}. Figures \ref{comp10} through \ref{comp10180} depict the FLI maps obtained for the previously listed values of the semi-major axis, by setting the initial time $t_0$ so that $M_S(t_0)+\om_S=0$, $\Om(t_0)=0$ and choosing $\om(t_0)$ equal to either $0^\circ$ or $180^\circ$. By making this choice $\sigma_{j,k}=0^\circ$ or $180^\circ$ for all the resonant angles. We remark that by choosing a different initial time or a nonzero value of $\Omega(t_0)$, the maps will be different since for some resonances we would be plotting the FLIs associated to a value of the resonant angle which does not correspond to an equilibrium point. In Figures \ref{comp10} and $\ref{comp1180}$ the value of the area-to-mass ratio is equal to $1$ m\textsuperscript{2}/kg , while in Figures \ref{comp100} and \ref{comp10180} it is equal to $10$ m\textsuperscript{2}/kg . Finally, Figures \ref{comp100om30} and \ref{comp10180om30} depict the case  with $\Om(t_0)=30^\circ$, for an object with area-to-mass equal to $10$ m\textsuperscript{2}/kg .

\bibliographystyle{plainnat}

\bibliography{cas-refs.bib}

\end{document}